\documentclass[a4paper,11pt]{article}
\pdfoutput=1 

\usepackage{jcappub} 

\usepackage[T1]{fontenc} 
\usepackage{caption}
\usepackage{subcaption}
\usepackage{graphicx}
\usepackage{mathrsfs}
\usepackage{amsmath}
\usepackage{amssymb}
\usepackage{mathtools}

\title{\boldmath Parker Bound and \\ Monopole Pair Production from Primordial Magnetic Fields}

\author[a,b,c,d]{Takeshi Kobayashi}
\author[a,b,c]{and Daniele Perri}

\affiliation[a]{SISSA, International School for Advanced Studies, \\via Bonomea 265, 34136 Trieste, Italy\\}

\affiliation[b]{INFN, Sezione di Trieste, \\via Valerio 2, 34127 Trieste, Italy\\}

\affiliation[c]{IFPU, Institute for Fundamental Physics of the Universe, \\via Beirut 2, 34014 Trieste, Italy\\}

\affiliation[d]{Kobayashi-Maskawa Institute for the Origin of Particles and the Universe, \\Nagoya University, Nagoya 464-8602, Japan\\}

\emailAdd{takeshi.kobayashi@sissa.it}
\emailAdd{dperri@sissa.it}

\abstract{We present new bounds on the cosmic abundance of magnetic monopoles based on the survival of primordial magnetic fields during the reheating and radiation-dominated epochs. The new bounds can be stronger than the conventional Parker bound from galactic magnetic fields, as well as bounds from direct searches. We also apply our bounds to monopoles produced by the primordial magnetic fields themselves through the Schwinger effect, and derive additional conditions for the survival of the primordial fields.}

\begin{document}
\maketitle
\flushbottom

\section{Introduction}
\label{intro}

Magnetic monopoles, albeit without any experimental evidence, are an inevitable prediction of theories of grand unification. They are point-like topological defects \cite{thooft,polyakov} which can be produced during phase transitions in the early universe \cite{Preskill,ZK}. The existence of magnetic monopoles is also related to the quantization of the electric charge via the Dirac quantization condition \cite{Dirac}. Their abundance in the universe today is subject to bounds from direct searches and from the requirement that they do not exceed the critical density of the universe \cite{pdg, Preskill, KolbTurner}.
Moreover, by noting that a population of monopoles would short out the magnetic fields inside galaxies, Parker obtained upper bounds on the flux of monopoles~\cite{TPB, Parker}.

Magnetic fields have been observed in the universe on different length scales, while their origin still remains unknown. Magnetic fields of $B \sim 10^{-5}~\mathrm{G}$ \cite{Widrow} have been observed within spiral galaxies. Recent gamma ray observations suggested the presence of magnetic fields even in intergalactic voids with strengths $B \gtrsim 10^{-15}~\mathrm{G}$ coherent on $\mathrm{Mpc}$ scales or larger \cite{TGFBGC,NV,DCRFCL}. The existence of such intergalactic magnetic fields gives strong indication that the fields have their origin in primordial magnetic fields produced in the early universe.

Magnetic fields accelerate the monopoles and the process of monopole acceleration extracts energy from the fields. The energy that the monopoles extract from the primordial magnetic field is consequently transferred to the primordial plasma through scattering processes with relativistic charged particles of the plasma \cite{VilShell}. With a monopole number density large enough, this can cause the disappearance of the field. 
Thus, from the survival of primordial magnetic fields until today, bounds similar to the Parker bound for galactic magnetic fields can be derived.
Such bounds from the primordial magnetic fields during the radiation-dominated epoch were analyzed by \cite{Vachaspati}.

Regarding the origin of the primordial magnetic fields, a class of scenarios that have been extensively studied invokes an explicit breaking of the Weyl invariance of the gauge field action, to excite magnetic fields during cosmic inflation~\cite{TW, Ratra} or in the subsequent epoch dominated by an oscillating inflaton~\cite{Kobayashi2}.\footnote{Cosmological phase transitions can also give rise to primordial magnetic fields~\cite{Vachaspati2,Cornwall}.}
A generic feature of these scenarios is that the magnetic fields are generated while the universe is cold, because after reheating completes the universe turns into a good conductor and the magnetic flux freezes in. 
Hence the interplay between primordial magnetic fields and monopoles may well have been important from times prior to radiation domination.

Although the magnetic fields in the universe today are rather weak, if they have a primordial origin, in the early universe they could have been extremely strong. Such strong magnetic fields can themselves produce monopole-antimonopole pairs through the magnetic dual of the Schwinger effect~\cite{Schwinger,AM,AAM}.
Even superheavy monopoles can thus be produced in primordial magnetic fields~\cite{PrimMag}.
The pair production in turn depletes energy from the magnetic fields, which, along with the subsequent acceleration of the produced monopoles, induce a self-screening of the fields.
The implications of monopole pair production in primordial magnetic fields were recently investigated in~\cite{PrimMag}, however, that work used a simplified treatment of the magnetic field dissipation by monopole acceleration. 
In particular, it focused on the acceleration right after the monopoles are pair produced, but did not take into account the integrated effect from monopole acceleration over the entire cosmological history.

In this work we present a comprehensive study of the Parker limit from primordial magnetic fields. 
We generalize the analysis of \cite{Vachaspati} and evaluate the effects of monopole acceleration in primordial magnetic fields throughout the post-inflation universe, starting from the reheating epoch when the universe is dominated by an oscillating inflaton. We show that, depending on the early cosmological history, the survival of primordial magnetic fields during reheating imposes bounds on the monopole abundance that are more stringent than the Parker limit from Galactic magnetic fields and the bound presented in \cite{Vachaspati}.

Even in the absence of any initial monopole population, strong magnetic fields in the early universe can Schwinger-produce monopole pairs. Hence we also apply our generic bounds to such pair-produced monopoles, in order to obtain the most conservative condition for the survival of primordial magnetic fields.
The bound we derive is comparable to those obtained in \cite{PrimMag} from considerations of the magnetic field screening by the Schwinger process, and the overproduction of monopoles.

This paper is organized as follows. In Section~\ref{general} we show the effects of the primordial magnetic fields and of the drag force of the plasma on the evolution of the monopole velocity from the time when the magnetic field is generated to the epoch of $e^+ e^-$ annihilation. In Section~\ref{numbBounds} we review the bound on the monopole flux from the persistence of the primordial magnetic fields during radiation domination, then we compute the corresponding bound for the reheating epoch.
In Section~\ref{prodField}, we apply our generic bounds to monopoles that are pair produced by the primordial magnetic fields.
We then conclude in Section~\ref{concl}. In Appendix~\ref{app1} we review the evolution of the universe after inflation.

Our analysis can be applied to both elementary and solitonic monopoles. The monopole magnetic charge is usually related to the electric charge as $g = 2 \pi n /e$, where $n \in \mathbb{N}$, and therefore $g$ is typically large (e.g. $g \simeq 21n$ for $e \simeq 0.30$). However, the possibility of milli-charged monopoles or monopoles with fractional charge has been analyzed recently \cite{BJ, HISY, HH}. In most of our discussions, we keep $g$ general without specifying its value to facilitate generalizations to various kinds of monopoles.

Within this paper we adopt Heaviside-Lorentz units, with $c = \hbar = k_B = 1$ and $M_{\mathrm{Pl}}$ corresponds to the reduced Planck mass $(8 \pi G)^{-1/2}$. We use Greek letters for spacetime indices and Latin letters when we mean only the three spatial components. We choose the metric tensor signature $(+---)$.

\section{Monopole dynamics in primordial magnetic fields}
\label{general}

Although the cosmological expansion history from Big Bang Nucleosynthesis onward is constrained by various observations \cite{Planck, pdg}, we have very few information on what happened before. For this work, we assume that at the time of the end of inflation, $t_{\mathrm{end}}$, the universe is initially dominated by an oscillating inflaton field, which decays perturbatively into radiation. At time $t_{\mathrm{dom}}$ the radiation component starts to dominate the universe, until matter domination begins at matter-radiation equality, at time $t_{\mathrm{eq}}$. During reheating, the cosmological plasma sourced by the inflaton decay is not necessary in thermal equilibrium, and then it is not possible to define a cosmic temperature. In any case, we assume the plasma to be in thermal equilibrium during both the reheating and radiation-dominated epochs. Imposing this assumption leads to a conservative bound, as we will explain later.
In Appendix~\ref{app1} we review the evolution of the Hubble rate and of the cosmic temperature during the reheating epoch and the subsequent radiation-dominated epoch.

Gamma ray observations suggest the existence of an intergalactic magnetic field $B_0 \gtrsim 10^{-15}~\mathrm{G}$ coherent on $\mathrm{Mpc}$ scales or larger \cite{TGFBGC,NV,DCRFCL} (the subscript ``$0$'' denotes quantities in the present universe). Throughout this paper we assume that this large-scale intergalactic magnetic field was produced in the early universe.
We expect that large-scale magnetic fields redshift as $B \propto a^{-2}$ in the absence of significant back-reaction from the monopoles or of any external source for the fields \cite{KobSlo}, where $a(t)$ is the scale factor.
Magnetic fields coherent on scales of $\mathrm{Mpc}$ have always been outside the Hubble horizon during the period that we are going to analyze.
Thus, the distance crossed by the monopoles during the period of interest is smaller than the correlation length of the magnetic fields. 
This allows us to consider the magnetic field to be effectively homogeneous.

We now describe the motion of monopoles in a homogeneous magnetic field with a friction force due to the primordial relativistic plasma.
Under these conditions, the general covariant form for the equation of motion of the monopoles is:
\begin{equation}
    m \mathrm{v}^{\nu} \nabla_{\nu} \mathrm{v}^{\mu} = m \left( \frac{d \mathrm{v}^{\mu}}{d \tau} + \Gamma^{\mu}_{\alpha \beta} \mathrm{v}^{\alpha} \mathrm{v}^{\beta} \right) = \mathcal{F}_{\mathrm{mag}}^{\mu} + \mathcal{F}_{\mathrm{p}}^{\mu} ,
\end{equation}
where $\nabla_{\nu}$ is the covariant derivative, $\mathcal{F}_{\mathrm{mag}}^{\mu}$ is the magnetic force responsible for the acceleration of the monopoles, $\mathcal{F}_{\mathrm{p}}^{\mu}$ is the drag force from the interaction with the primordial plasma, $\Gamma^{\mu}_{\alpha \beta}$ are the Christoffel symbols of the metric, $m$ is the monopole mass, $\mathrm{v}^{\mu}$ is the four-velocity of the monopole, with $\mathrm{v}_{\mu} \mathrm{v}^{\mu} = 1$, and $\tau$ its proper time.
The magnetic force can be expressed through the four-vector \cite{Jackson}:
\begin{equation}
    \mathcal{F}_{\mathrm{mag}}^{\mu} = g \Tilde{F}^{\mu \nu} \mathrm{v}_{\nu} ,
\end{equation}
where $\Tilde{F}^{\mu \nu} = \frac{1}{2} \epsilon^{\mu \nu \alpha \beta} F_{\alpha \beta}$ is the dual electromagnetic tensor and $\epsilon^{\mu \nu \alpha \beta}$ is a totally antisymmetric pseudotensor normalized as $|\epsilon^{\mu \nu \alpha \beta}| = 1 / \sqrt{- \mathrm{det}(g_{\rho \sigma})}$, if $\mu$, $\nu$, $\alpha$ and $\beta$ are all distinct. 
Without loss of generality, we take the magnetic charge of the monopole $g$ to be positive.

We limit our analysis to times before $e^+ e^-$ annihilation, i.e. $T \gtrsim 1~\mathrm{MeV}$, when the cosmological plasma consists of relativistic charged particles.
Monopoles interact with the plasma through elastic scattering $M + x^{\pm} \rightarrow M + x^{\pm}$, where $x^{\pm}$ is a generic charged particle of the Standard Model and beyond\footnote{The details of the calculation and phenomenological aspects of the effective operator of the interaction can be found in \cite{MaMi, AlMa}.}. The result is an effective drag force acting on the monopoles. We adopt the covariant form of the drag force shown in \cite{VilShell}:
\begin{equation}
    \mathcal{F}_{\mathrm{p}}^{\mu} = f_{\mathrm{p}}~\mathrm{h}(v_{\mathrm{rel}})~(u \cdot \mathrm{v})^{-1}~[u^{\mu} - (u \cdot \mathrm{v}) \mathrm{v}^{\mu}],
\end{equation}
where $u^{\mu}$ is the mean four-velocity of the particles in the plasma and $v_{\mathrm{rel}} = (1 - (u \cdot \mathrm{v})^{-2})^{1/2}$ is the velocity of the monopole in the rest frame of the plasma. Here $\mathrm{h} (w)$ is a slowly-varying function with $\mathrm{h}(0) = 1$ and $\mathrm{h} (1) = 3/2$:
\begin{equation}
    \mathrm{h} (w) = \frac{3}{2 w^2} \left[ 1 + \frac{1 - w^2}{2 w} \ln{\left(\frac{1-w}{1+w} \right)} \right] .
\end{equation}
For simplicity we fix $\mathrm{h} (v_{\mathrm{rel}})$ to unity in the following analyses. For relativistic scatterers that are in thermal equilibrium, $f_{\mathrm{p}}$ can be expressed as\footnote{The expression of $f_{\mathrm{p}}$ is different by a factor $16 \pi^2$ from that in \cite{Vachaspati}, which used CGS units instead of Heaviside-Lorentz.}:
\begin{equation}
\label{effep}
    f_{\mathrm{p}} \sim \frac{e^2 g^2 \mathcal{N}_c}{16 \pi^2} T^2 ,
\end{equation}
with $\mathcal{N}_c$ the number of relativistic and electrically charged degrees of freedom in thermal equilibrium including also the contributions of the spin and the charge of the scatterers. In this paper, we always assume for the electric charge a value $e = 0.30$.
\footnote{Before the electroweak phase transition, the monopoles and primordial magnetic fields are those of the hypercharge U(1), and thus quantities such as the charge are modified by a number of order unity that depends on the Weinberg angle. We ignore this effect, as well as the running of the parameters; this treatment should be good enough for the order-of-magnitude calculations in this paper.}
Due to the drag force, the energy in the magnetic fields that is used to accelerate the monopoles eventually gets dissipated into the thermal plasma\footnote{Depending on the amount of the magnetic field energy that is dissipated, the plasma experiences an additional reheating that can have non-negligible effects on the evolution of the universe.}.

We now consider a Friedmann-Robertson-Walker (FRW) background spacetime $ds^2 = dt^2 - a^2 dx^i dx^i$, assuming sum over repeated spatial indices irrespective of their positions, and suppose the plasma to be at rest in the coordinate system $(t, x^i)$, i.e. $u^{\mu} = (1,0,0,0)$. In this reference frame, the velocity of the monopoles can be expressed as $\mathrm{v}^{\mu} = (\gamma, \gamma v^i/a)$, with $\gamma = \frac{1}{\sqrt{1 - v^2}}$, $v^i = a (dx^i/dt)$, and $v = (v^i v^i)^{1/2}$ the modulus of the three-velocity. 

In the absence of magnetic fields, the motion of the monopoles can be described as a Brownian motion within the plasma \cite{VilShell}. Consequently, the monopoles present thermal velocities $v_T \sim \left( T/m \right)^{1/2}$, with zero mean velocity after taking an average over the three directions. In the early univese, the thermal velocity can be larger than the mean velocity induced by a primordial magnetic field. However in this work we ignore the thermal velocity, assuming that it does not leave any coherent effects on large scales.
The mechanisms of monopole production that we assume for our analysis are not able to give the monopoles a significant mean velocity upon production. In addition, any initial mean velocity of the monopoles decays away due to the drag forces.
This allows us to assume for the monopoles a zero mean velocity when the magnetic fields are generated, simplifying the analysis.

The magnetic field vector is $B^{\mu} = \Tilde{F}^{\mu 0}$, with $B^{\mu} B_{\mu} = - B^2$ and $B$ the amplitude of the magnetic field. Choosing the $x^3$-axis along the direction of the magnetic field (namely, $B^3 = B/a$ and $B^1 = B^2 = 0$), we can ignore the monopole velocity along the other directions, i.e. $v^3 = v$ and $v^1 = v^2 = 0$. Under these assumptions, the motion of the monopoles can be described by the equation for the average velocity:
\begin{equation}
\label{eqVelocity}
    m \frac{d}{dt} ( \gamma v) = gB - \left( f_{\mathrm{p}} + m H \gamma \right) v ,
\end{equation}
where $H(t) = \dot{a}/a$ is the Hubble rate, and an overdot denotes a time derivative. The contribution of the universe expansion can be seen as an additional frictional term proportional to the Hubble rate. 

We do not specify the detailed mechanism for the generation of the primordial magnetic fields. According to the models proposed in the literature, the magnetic fields can be generated during inflation \cite{TW, Ratra}, after inflation when the universe is dominated by an oscillating inflaton \cite{Kobayashi2}, or at cosmological phase transitions \cite{Vachaspati2, Cornwall}.

We define $t_i$ as the moment when the generation of the magnetic fields has concluded and the fields start to redshift freely with the expansion of the universe (the subscript ``$i$'' denotes quantities at the end of magnetic field generation).
In this work we assume the primordial magnetic field to be suddenly switched on at time $t_i$. 
This corresponds to considering only times after the end of the process of the magnetic field generation. 
Although it is possible to obtain further constraints taking into account also the time interval during the production of the magnetic fields, we leave this for future analysis.
For this work we assume that $t_i$ is at the end of inflation or in the subsequent epochs. In other words, we consider the Hubble rate during inflation to be no smaller than that at the magnetic field generation, i.e. $H_{\mathrm{inf}} \geq H_i$.

At time $t_i$, monopoles have mean velocity equal to zero. (Here we are tacitly assuming that the monopoles are present when the magnetic fields are switched on. For monopoles produced afterwards the `initial time' should be taken as the time when the monopoles are produced.) Considering time intervals $t - t_i$ shorter than the time scales of the frictional forces (which will be specified below), the effects of the universe expansion and of the plasma can be ignored and the product of the velocity and gamma factor can be expressed as:
\begin{equation}
    \gamma v \simeq \frac{gB_i}{m}(t-t_i) .
\end{equation}
Thus, monopoles can be freely accelerated to relativistic or non-relativistic velocities depending on the intensity of the fields and on their mass. At later times, the frictional terms become important, and the velocity of the monopoles starts to decrease. Depending on the temperature, one of the frictional terms eventually dominates over the other, giving rise to different behaviors of the velocity. We analyze the velocity evolution for two different regimes: during radiation domination ($t>t_{\mathrm{dom}}$), and during the reheating epoch ($t<t_{\mathrm{dom}}$).

\subsection{During radiation domination}
\label{radDomVel}

During radiation domination, the Hubble rate and the cosmic temperature redshift as $H \propto a^{-2}$ and $T \propto a^{-1}$, up to the time variation of the number of relativistic degrees of freedom, $g_{*(s)}$. For $t > t_{\mathrm{dom}}$, let us for the moment assume that the Hubble friction on the monopoles is negligible. We also assume that the monopoles move at non-relativistic velocities because of the interaction with the plasma.
Under these assumptions, the equation of motion of the monopoles can be rewritten as:
\begin{equation}
     m \dot{v} = gB - f_{\mathrm{p}} v .
\end{equation}
Neglecting the time variation of $B$, $f_{\mathrm{p}}$ and $H$, the general solution of the equation is $v = C \exp \left( - (f_{\mathrm{p}}/m) t \right) + v_{\mathrm{p}}$, where $C$ is a constant that depends on the initial conditions and $v_{\mathrm{p}}$ is the terminal velocity:
\begin{equation}
\label{vTermPlasma}
    v_{\mathrm{p}} = \frac{g B}{f_{\mathrm{p}}} \sim \frac{16 \pi^2 B}{e^2 g \mathcal{N}_c T^2} .
\end{equation}
The characteristic time necessary for the monopoles to feel the effects of the interaction with the particles of the plasma can then be defined as $\Delta t_{\mathrm{p}} \sim m/f_{\mathrm{p}} \simeq \left(16 \pi^2 m \right) / \left( e^2 g^2 \mathcal{N}_c T^2 \right)$ \cite{Vachaspati}. 
After a time $\Delta t_{\mathrm{p}}$, the monopoles approach the terminal velocity $v_{\mathrm{p}}$.

As long as we consider magnetic fields of order $10^{-15}~\mathrm{G}$ today, they cannot have dominated the energy density of the universe during radiation domination, i.e. $B \ll T^2$. Thus, for $g \sim 2 \pi / e$ the expression in Eq.~\eqref{vTermPlasma} gives $v_{\mathrm{p}} \ll 1$. This justifies our use of non-relativistic equations.

Comparing the timescale $\Delta t_{\mathrm{p}}$ with the Hubble time $\Delta t_{H} \sim 1/H \sim M_{\mathrm{Pl}}/ (g_*^{1/2} T^2)$, we observe that for a magnetic charge of $g = 2 \pi / e$ the effect of the expansion of the universe can be neglected for masses:
\begin{equation}
\label{subplanck}
    m < \frac{M_{\mathrm{Pl}} \mathcal{N}_c}{g_*^{1/2}} .
\end{equation}
Limiting our analysis to sub-planckian values for the masses of the monopoles, this condition is always satisfied. Therefore, this justifies our assumption of neglecting the Hubble friction during radiation domination.

Using $T_0 \sim 10^{-4}~\mathrm{eV}$ and $B_0 \sim 10^{-15}~\mathrm{G} \simeq 2 \cdot 10^{-17}~\mathrm{eV^2}$, the terminal velocity at $T \sim 1~\mathrm{MeV}$ (at which time $\mathcal{N}_c \sim g_{*s} \simeq 10.75$) is estimated as $v_{\mathrm{p}} \sim 10^{-8}$ for $g \sim 2\pi / e$.

\subsection{Before radiation domination}
\label{2.1.2}

Within this work, for simplicity we assume the total number of relativistic degrees of freedom $g_{*(s)}$, as well as the number of charged relativistic degrees of freedom $\mathcal{N}_c$, to stay constant for $t < t_{\mathrm{dom}}$.
During the reheating epoch, assuming that the plasma is in thermal equilibrium, then $H \propto a^{-3/2}$ and $T \propto a^{-3/8}$ (see Appendix~\ref{app1} for the computation). Consequently, the Hubble friction can play an important role in the monopole dynamics, and moreover the monopoles can move with relativistic velocities.

Thus, let us compare $f_{\mathrm{p}}$ and $m H \gamma$  in the equation of motion in Eq.~\eqref{eqVelocity}, to see which of the friction terms dominates during the reheating epoch.
We introduce the ratio $r(t) = \rho_{\mathrm{rad}} (t) / \rho_{\mathrm{tot}} (t)$, where $\rho_{\mathrm{tot}}$ is the total energy density of the universe and $\rho_{\mathrm{rad}}$ the energy density in radiation, with $r \leq 1/2$. The value of $r$ decreases going back in time. The Hubble rate then can be expressed as:
\begin{equation}
    H \sim \frac{g_*^{1/2} T^2}{r^{1/2} M_{\mathrm{Pl}}} \gtrsim \frac{g_*^{1/2} T^2}{M_{\mathrm{Pl}}} .
\end{equation}

If the monopoles move at non-relativistic velocities and assuming $g \sim 2 \pi / e$, in order for $m H \gamma$ to be smaller than $f_{\mathrm{p}}$, the radiation fraction needs to satisfy:
\begin{equation}
\label{req}
    r \gtrsim \left( \frac{g_*^{1/2} m}{\mathcal{N}_c M_{\mathrm{Pl}}} \right)^2 .
\end{equation}
For example, with a mass $m \simeq 10^{16}~\mathrm{GeV}$ and with $g_{*} \sim \mathcal{N}_c \simeq 100$, the condition in Eq.~\eqref{req} can be read as $r \gtrsim 10^{-7}$. If the magnetic fields are generated sufficiently in the past, going back in time eventually the condition in Eq.~\eqref{req} breaks down. 
This signals $f_p < m H \gamma$, namely, the Hubble friction dominates, and the equation of motion of the monopoles can be approximately written as:
\begin{equation}
     m \frac{d}{dt} ( \gamma v) = gB - m H \gamma v .
\end{equation}
The terminal velocity can be estimated by equating the terms in the right-hand side as,\footnote{The time scales for $v$ to achieve $v_{\mathrm{H}}$, and for the redshifting of $B$ and $H$, are all of order the Hubble time. Hence $d (\gamma v) / dt$ actually does not vanish and the terminal velocity is $( \gamma v )_{\mathrm{H}} = 2 g B/m H$ (see Eq.~(A7) in \cite{PrimMag} for the derivation). However we will omit the factor~2 since we are interested in order-of-magnitude estimates.}
\begin{equation}
\label{vTermHubble}
    \left( \gamma v \right)_{\mathrm{H}} \sim \frac{g B}{m H} .
\end{equation}
This can also take relativistic values, unlike the terminal velocity in Eq.~(\ref{vTermPlasma}) due to the plasma friction.

\begin{figure}[!t]
  \centering
  \begin{subfigure}[b]{0.72\textwidth}
    \centering
    \includegraphics[width=\textwidth]{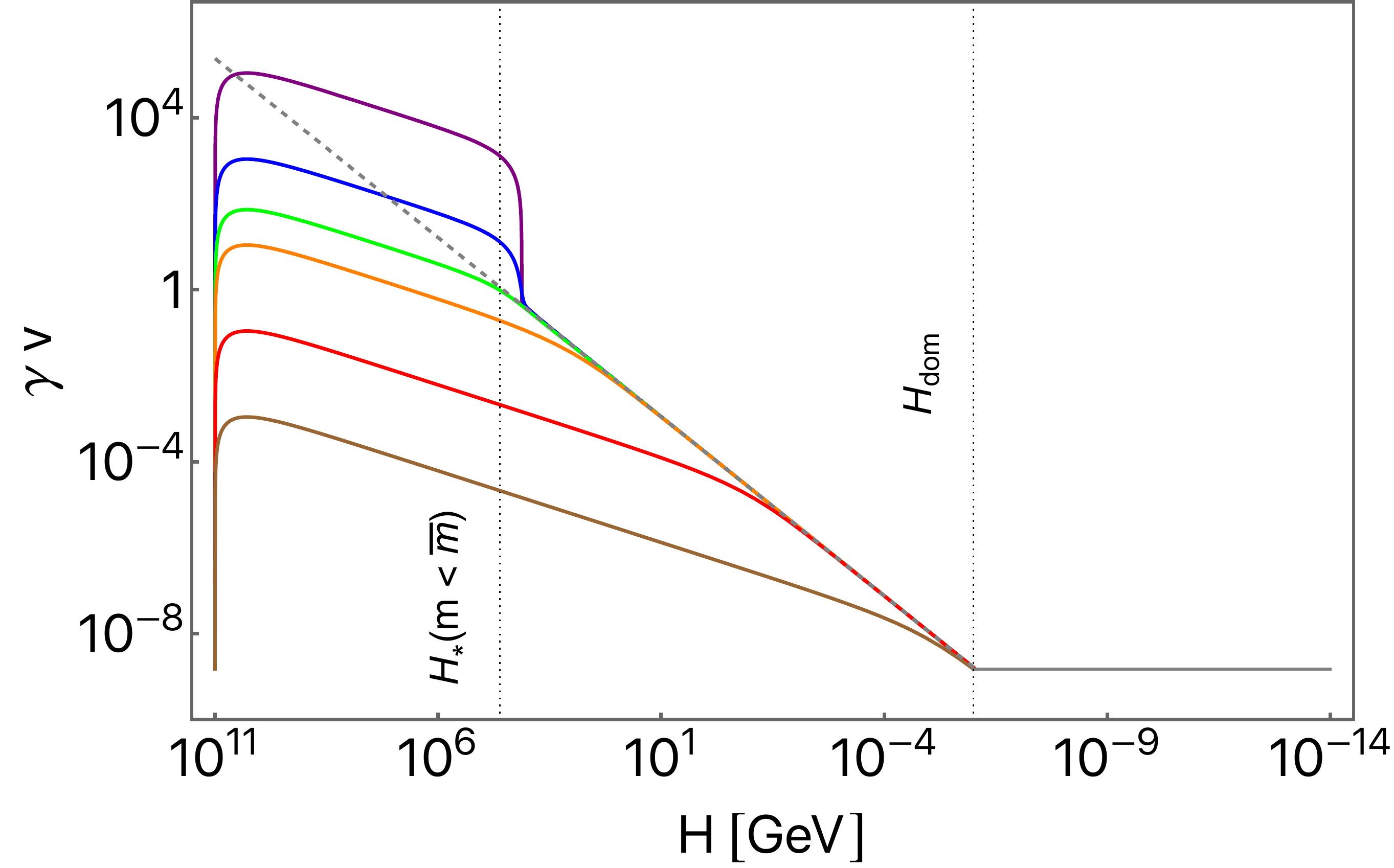}
    \caption{}
    \label{velocity}
  \end{subfigure}
  \hfill
  \begin{subfigure}[b]{0.7\textwidth}
    \centering
    \includegraphics[width=\textwidth]{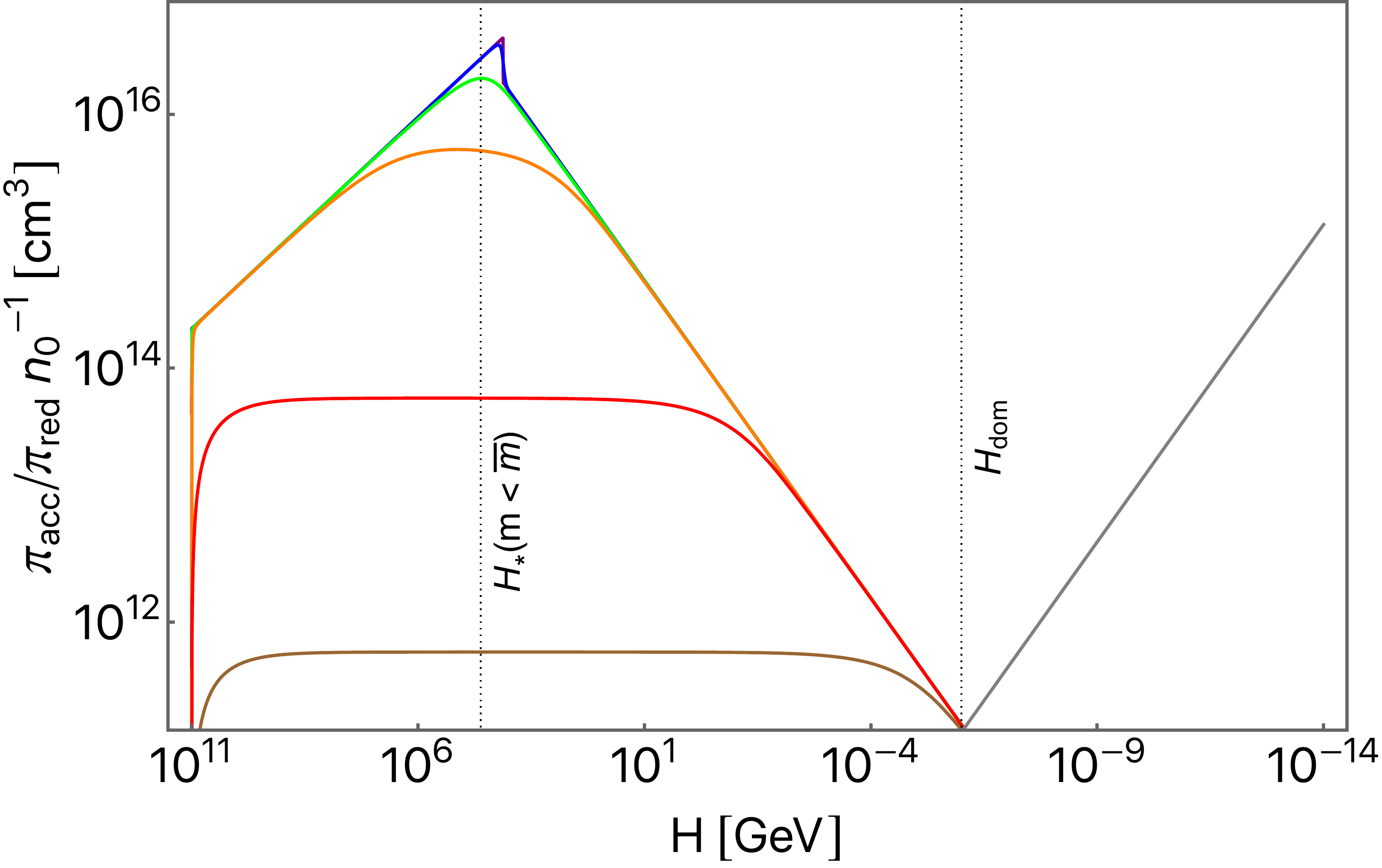}
    \caption{}
    \label{ratio}
  \end{subfigure}
  \caption{Evolution of the monopole velocity in primordial magnetic fields (top) and of the normalized damping rate of the magnetic fields (bottom) for different values of the monopole mass (from bottom to top, brown: $m = 10^{19}~\mathrm{GeV}$; red: $m = 10^{17}~\mathrm{GeV}$; orange: $m = 10^{15}~\mathrm{GeV}$; green: $m = 10^{14}~\mathrm{GeV}$; blue: $m = 10^{13}~\mathrm{GeV}$; purple: $m = 10^{11}~\mathrm{GeV}$). The expression in Eq.~\eqref{vTermPlasma} for the terminal velocity set by the friction with the thermal plasma is also shown in dashed line in the top plot. Here $H_{\mathrm{dom}} = 10^{-6}~\mathrm{GeV}$, $B_0 = 10^{-15}~\mathrm{G}$, $g = 2 \pi /e$, $g_{*} = \mathcal{N}_c = 100$ and we use $H_i = 10^{11}~\mathrm{GeV}$ as the starting point of the evolution. The value of $H_* \sim 10^{5}~\mathrm{GeV}$ for monopole masses smaller than $\bar{m}$ is also shown in the plots (see the text for details).}
  \label{VelRatio}
\end{figure}

In Figure~\ref{velocity} we show the time evolution for $\gamma v$, by numerically solving the equation of motion Eq.~\eqref{eqVelocity}. The results are shown for $H(t) < H_i$, and for different values of the monopole mass. For $H(t) > H_{\mathrm{dom}}$, each value of the mass is associated to a differently colored solid curve (from bottom to top, brown: $m = 10^{19}~\mathrm{GeV}$; red: $m = 10^{17}~\mathrm{GeV}$; orange: $m = 10^{15}~\mathrm{GeV}$; green: $m = 10^{14}~\mathrm{GeV}$; blue: $m = 10^{13}~\mathrm{GeV}$; purple: $m = 10^{11}~\mathrm{GeV}$). 
The dashed line in the regime $H(t) > H_{\mathrm{dom}}$ shows $v_{\mathrm{p}}$ given in Eq.~\eqref{vTermPlasma}, which corresponds to the terminal velocity set by the plasma when $v_{\mathrm{p}} \ll 1$. 
For $H(t) < H_{\mathrm{dom}}$ the velocity is constant and independent of the mass of the monopoles, hence it is represented by a single solid horizontal grey line.
For the computation we assume $g = 2 \pi /e$, $B_0 = 10^{-15}~\mathrm{G}$ and $g_{*} = \mathcal{N}_c = 100$ throughout.
We use the results presented in Appendix~\ref{app1} for setting the time dependence of $a$, $T$ and $B$. 
We start the computation at $H_i = 10^{11}~\mathrm{GeV}$, with an initial condition $v(t_i) = 0$. Moreover, we choose the cosmic temperature when radiation domination takes over as $T_{\mathrm{dom}} = 10^6~\mathrm{GeV}$, which corresponds to the Hubble rate $H_{\mathrm{dom}} \simeq 10^{-6}~\mathrm{GeV}$ (the subscript ``$\mathrm{dom}$'' denotes quantities computed at time $t_{\mathrm{dom}}$).

Independently from the initial condition, the monopole velocity rapidly falls into one of the attractor solutions, $v_{\mathrm{p}}$ in Eq.~\eqref{vTermPlasma} and $v_{\mathrm{H}}$ in Eq.~\eqref{vTermHubble}.
The evolution of the velocity for $H(t) < H_i$ is hence independent of the value of $H_i$. 
However, the choice of $H_i$ determines how far back in time can one go with the attractor solutions.
With a sufficiently large $H_i$, the Hubble friction initially dominates over the friction from the plasma, yielding $(\gamma v)_{\mathrm{H}}$ which redshifts as $H^{1/3}$.
During reheating the fraction of energy density in the radiation component increases with time and at some point Eq.~\eqref{req} starts to be satisfied. This is the signal that the velocity begins to be controlled by the friction from the plasma, cf. Eq.~\eqref{vTermPlasma}, and then the velocity decreases as $H^{5/6}$. For masses satisfying the condition in Eq.~\eqref{subplanck}, monopoles achieve $v_{\mathrm{p}}$ before radiation domination and the value of the velocity at $T=T_{\mathrm{dom}}$ is independent of the mass.

Of crucial relevance is the time $t_*$ of the transition between the domination of the Hubble friction term and that of the friction term by the primordial plasma: 
\begin{equation}
\label{fstarcondition}
    f_{\mathrm{p}, *} = m H_* \gamma_* ~,
\end{equation}
where the subscript ``$*$'' stands for quantities computed at time $t_*$. For $t < t_*$ the monopoles move at the terminal velocity set by the expansion of the universe shown in Eq.~\eqref{vTermHubble}. For $t > t_*$ the frictional term due to the interaction with the plasma dominates the evolution and the velocity of the monopoles can be expressed through Eq.~\eqref{vTermPlasma}.
Rewriting Eq.~\eqref{fstarcondition} as an expression for the velocity of the monopoles $v_*$, we get:
\begin{equation}
    v_*^2 = 1 - \left( \frac{m H_*}{f_{\mathrm{p}, *}} \right)^2 .
\end{equation}
When $v > v_*$ the motion of the monopole is set by the Hubble friction term, while for $v < v_*$ it is dominated by the friction force of the plasma. 
We can obtain the Hubble rate at the transition by using Eq.~\eqref{vTermHubble} and setting $v_* \simeq v_{H, *}$ as:
\begin{equation}
\label{conditionStar}
    \frac{m^2 H_*^2}{f_{\mathrm{p}, *}^2} + \frac{g^2 B_*^2}{f_{\mathrm{p}, *}^2} \simeq 1 .
\end{equation}
Considering that $f_{\mathrm{p}} \propto H^{1/2}$ and $B \propto H^{4/3}$ during reheating, we can rewrite Eq.~\eqref{conditionStar} in terms of quantitites at $t_\mathrm{dom}$ as:
\begin{equation}
\label{eqHstar}
    \alpha  \left( \frac{H_*}{H_{\mathrm{dom}}} \right)^{5/3} + \beta (m)  \left( \frac{H_*}{H_{\mathrm{dom}}} \right) \simeq 1 ,
\end{equation}
where we define:
\begin{subequations}
\begin{equation}
    \alpha = \left( \frac{g B_{\mathrm{dom}}}{f_{\mathrm{p},\mathrm{dom}}} \right)^{2} ,
\end{equation}
\begin{equation}
    \beta (m) = \left( \frac{m H_{\mathrm{dom}}}{f_{\mathrm{p},\mathrm{dom}}} \right)^{2} .
\end{equation}
\end{subequations}
Depending on the value of the monopole mass, one of the two terms on the left-hand side of Eq.~\eqref{eqHstar} is larger than the other. 
The two terms are always positive and they are of the same order only if both of them are of order unity, i.e. $\alpha  (H_* / H_{\mathrm{dom}})^{5/3} \sim \beta (\bar{m})  (H_* / H_{\mathrm{dom}}) \sim 1$, where we define $\bar{m}$ as the mass for which the two terms are comparable. 
From these considerations we obtain the relation for $\bar{m}$:
\begin{equation}
\label{mbar}
    \alpha  \beta(\bar{m})^{-5/3} \sim 1 .
\end{equation}
Substituting the definition for $\alpha$ and $\beta$, we get the explicit expression for $\bar{m}$:
\begin{equation}
    \bar{m} \sim \frac{\left( g^{3} B_{\mathrm{dom}}^{3} f_{\mathrm{p},\mathrm{dom}}^{2} \right)^{1/5}} {H_{\mathrm{dom}}} .
\end{equation}
Using Eq.~\eqref{aTdom} in Appendix~\ref{app1} to rewrite the expressions for $f_{\mathrm{p},\mathrm{dom}}$, $B_{\mathrm{dom}}$ and $H_{\mathrm{dom}}$ in terms of their values at the present time, we obtain:
\begin{equation}
\label{mStar}
    \bar{m} \simeq 10^{14}~\mathrm{GeV}  \left( \frac{B_0}{10^{-15}~\mathrm{G}} \right)^{3/5} \left( \frac{g}{10} \right)^{7/5} \left( \frac{\mathcal{N}_{c, \mathrm{dom}}}{100} \right)^{2/5} .
\end{equation}
Assuming $B_0 \sim 10^{-15}~\mathrm{G}$, $\mathcal{N}_c \sim 100$ and $g \sim 2\pi / e \sim 10$, we get $\bar{m} \simeq 10^{14}~\mathrm{GeV}$. Thus, in Figure \ref{velocity} the green curve corresponds to the evolution of the velocity of monopoles with a mass $m = \bar{m}$.

For $m \ll \bar{m}$ the left-hand side of Eq.~\eqref{eqHstar} is dominated by the first term and the expression for $H_*$ is independent of the monopole mass:
\begin{equation}
    \frac{H_*}{H_{\mathrm{dom}}} \simeq \frac{1}{\alpha^{3/5}} = \left( \frac{f_{\mathrm{p},\mathrm{dom}}}{g B_{\mathrm{dom}}} \right)^{6/5} ,
\end{equation} 
while for $m \gg \bar{m}$ the second term dominates and: 
\begin{equation}
    \frac{H_*}{H_{\mathrm{dom}}} \simeq \frac{1}{\beta (m)} = \left( \frac{f_{\mathrm{p},\mathrm{dom}}}{m H_{\mathrm{dom}}} \right)^{2} .
\end{equation}
Using Eq.~\eqref{aTdom}, the expressions for $H_*$ can be further rewritten as:
\begin{equation}
\label{HstarBoth}
    H_* \simeq
    \begin{dcases}
        10^{4}~\mathrm{GeV}  \left( \frac{g}{10}\right)^{6/5} \left( \frac{ \mathcal{N}_{c, \mathrm{dom}}}{100} \right)^{6/5} \left( \frac{T_{\mathrm{dom}}}{10^{6}~\mathrm{GeV}} \right)^2 \left( \frac{10^{-15}~\mathrm{G}}{B_0} \right)^{6/5} &,~m \ll \bar{m} , \\
        10^{4}~\mathrm{GeV}  \left( \frac{g}{10} \right)^4 \left( \frac{ \mathcal{N}_{c, \mathrm{dom}}}{100} \right)^2 \left( \frac{T_{\mathrm{dom}}}{10^{6}~\mathrm{GeV}} \right)^2 \left( \frac{10^{14}~\mathrm{GeV}}{m} \right)^2 &,~m \gg \bar{m} .
    \end{dcases}
\end{equation}
In Figure~\ref{velocity} we plot in vertical line the value of $H_*$ in the limit $m \ll \bar{m}$, i.e. the first line of Eq.~\eqref{HstarBoth}.

For $m \ll \bar{m}$, the monopole velocity is always relativistic while it is on the Hubble-friction branch, $v = v_{\mathrm{H}}$; this is seen for the purple and blue curves in the plot. On the other hand, for $m \gg \bar{m}$, the monopoles become non-relativistic before switching to the plasma-friction branch, $v = v_{\mathrm{p}}$, as it is seen for the brown, red, and orange curves.

Let us estimate the time it takes for monopoles with $m \ll \bar{m}$ to jump from a relativistic $v_{\mathrm{H}}$ branch to a non-relativistic $v_{\mathrm{p}}$ branch. For times $t \lesssim t_*$ we can consider $v \simeq 1$ and the equation of motion for the monopoles can be rewritten in terms of the relativistic factor $\gamma$:
\begin{equation}
    \dot{\gamma} = \frac{gB - f_{\mathrm{p}}}{m} - H \gamma .
\end{equation}
Using $B \propto a^{-2}$, $f_{\mathrm{p}} \propto a^{-3/4}$, and $H \propto a^{-3/2}$, one can check that this equation has a solution,
\begin{equation}
 \gamma = 2 \frac{g B}{m H} - \frac{4}{7} \frac{f_{\mathrm{p}}}{m H}
= \gamma_* \left[ \frac{11}{7} \left(\frac{a}{a_*}\right)^{-1/2} - \frac{4}{7} \left( \frac{a}{a_*} \right)^{3/4} \right],
\label{eq:2.27}
\end{equation}
which asymptotes to $\gamma \simeq 2 g B / m H$ in the past. Upon moving to the far right-hand side, we used Eq.~(\ref{fstarcondition}). For $\gamma_* \gg 1$, the $\gamma$ factor approaches unity at $a = (11/4)^{4/5} a_* \simeq 2.2 a_*$, which is obtained by equating the terms in the square brackets in Eq.~(\ref{eq:2.27}).
Thus, the jump from an ultra-relativistic $v_{\mathrm{H}}$ to a non-relativisitc $v_{\mathrm{p}}$ happens with a time scale of $1/H_*$, as shown in the figure for the purple and blue curves.

Before closing this section, we should also remark that, as one goes back in time in the reheating epoch, the energy density in primordial magnetic fields grows relative to the total density as $\rho_{B} / \rho_{\mathrm{tot}} \propto a^{-1}$. 
Hence for primordial magnetic fields generated at the end of inflation or during reheating, i.e. $t_{\mathrm{end}} \leq t_i < t_{\mathrm{dom}}$, 
requiring that they have never dominated the universe constrains the time when magnetic field generation happened. The constraint is written using (\ref{aTdom}) and (\ref{aidom}) as,
\begin{equation}
 H_i \lesssim 10^{22}\, \mathrm{GeV}
\left( \frac{T_{\mathrm{dom}}}{10^6\, \mathrm{GeV}} \right)^2
\left(\frac{10^{-15}\, \mathrm{G}}{B_0}\right)^3.
\label{eq:2.28}
\end{equation}
With a reasonable choice of parameters (such as $g \sim 10$, $B_0 \sim 10^{-15}~\mathrm{G}$, and $\mathcal{N}_{c, \mathrm{dom}} \sim 100$), one sees that $H_*$ in Eq.~(\ref{HstarBoth}) is well below this limit on $H_i$, independently of $T_{\mathrm{dom}}$ and $m$.

\section{Bounds from the survival of primordial magnetic fields}
\label{numbBounds}

In the previous section we treated the primordial magnetic field as a background, however the magnetic fields themselves lose energy as they accelerate the monopoles. Here we study the energy depletion of the magnetic field, and derive conditions for the primordial magnetic field to survive until today. 
The kind of bounds analyzed in this section are analogous to the Parker bound \cite{TPB} but with respect to large-scale magnetic fields in the early universe, instead of galactic magnetic fields in the recent universe\footnote{See also \cite{YangBai} for an extension of the Parker bound using Andromeda galaxy.}.
The bound during radiation domination has already been analyzed by \cite{Vachaspati}. In this section we review the bound for times $t>t_{\mathrm{dom}}$ and we extend the analysis for $t<t_{\mathrm{dom}}$.

The magnetic fluid is described by the barotropic equation of state $P_{\mathrm{B}}/\rho_{\mathrm{B}} = 1/3$, where $P_B$ is the pressure of the magnetic fluid and $\rho_{\mathrm{B}} = B^2/2$ is the physical energy density. This corresponds to assuming that there are no external sources for the magnetic field for $t > t_i$. Under the hypothesis of a spatially homogeneous magnetic field, the evolution of the energy density $\rho_B$ within a FRW background, taking into account the effects of the monopole acceleration, is described by:
\begin{equation}
\label{enDensity}
    d\left[\rho_{\mathrm{B}}(t) a(t)^{3}\right]=-P_{\mathrm{B}}(t) d\left[a(t)^{3}\right]- 2 g B(t) d t \int_{-\infty}^{t} d t^{\prime} a\left(t^{\prime}\right)^{3} \Gamma\left(t^{\prime}\right) v\left(t^{\prime}, t\right),
\end{equation}
where $\Gamma (t)$ is the production rate for either monopoles or antimonopoles at time $t$. The second term on the right-hand side of the equation denotes the loss of energy due to accelerating the population of monopoles and antimonopoles produced from the infinite past to time $t$. All the possible production mechanisms that we take into account, i.e. monopole pairs produced by the magnetic fields, thermal production and production during phase transitions, cannot produce an asymmetry on the number of monopoles with respect to that of antimonopoles. Therefore, we always consider the monopole and antimonopole number densities to be equal and there is a factor 2 in the second term of Eq.~\eqref{enDensity}.
Here $a(t')^3 \Gamma (t') dt'$ represents the comoving number density of monopoles produced between $t'$ and $t' +dt'$, $v(t',t)$ corresponds to the velocity in the direction of the magnetic field at time $t$ of monopoles produced at $t'$ (with $t \geq t'$).
For the antimonopoles, we choose a charge of $-g$ and a velocity $-v(t',t)$.\footnote{Primordial magnetic fields can lose their energy also by pair-producing monopoles through the Schwinger effect. This effect can be taken into account by adding a term $-2 m \Gamma_{\mathrm{sw}}(t) a(t)^{3} d t$ on the right-hand side of Eq.~\eqref{enDensity}, where $\Gamma_{\mathrm{sw}}$ is the rate of monopole pair production through the Schwinger effect. Back-reaction of pair production on the magnetic field energy density has been studied in \cite{PrimMag}.}

The monopoles can be produced at a phase transition \cite{PreskillRef, Preskill}, through a thermal process \cite{Thermal1, Thermal2}, or by the magnetic field itself via the Schwinger effect. In this section we keep the discussion general and do not specify the production mechanism. The particular case of the Schwinger pair production will be the topic of Section~\ref{prodField}.

It has been shown that the monopole-antimonopole annihilation is relevant only if the monopole abundance is large enough to overdominate the universe \cite{Preskill, ZK}. Thus, we assume monopole-antimonopole annihilation to be negligible.

Eq.~\eqref{enDensity} can be then rewritten as:
\begin{equation}
\label{piEq}
    \frac{\dot{\rho}_{\mathrm{B}}}{\rho_{\mathrm{B}}}=-\Pi_{\mathrm{red}} - \Pi_{\mathrm{acc}},
\end{equation}
where $\Pi_{\mathrm{red}}$ and $\Pi_{\mathrm{acc}}$ are the dissipation rates of the magnetic field energy due to redshifting and monopole acceleration:
\begin{subequations} \label{PiAll}
\begin{equation}
\label{PiRed}
\Pi_{\mathrm{red}}(t) = 4 H(t),
\end{equation}
\begin{equation}
\Pi_{\mathrm{acc}}(t) = \frac{4 g}{a(t)^{3} B(t)} \int_{-\infty}^{t} d t^{\prime} a\left(t^{\prime}\right)^{3} \Gamma\left(t^{\prime}\right) v\left(t^{\prime}, t\right).
\end{equation}
\end{subequations}
Once the expression for the monopole velocity and for the total production rate is given, the evolution of the magnetic field energy density can be derived by solving Eq.~\eqref{piEq}. 

In the previous section we have shown that the memory of the monopole velocity at the time when the magnetic field is switched on is quickly lost. 
Hence, we assume that the monopoles have a uniform velocity independent of when they were produced, i.e. $v(t', t) = v(t)$. Under this assumption, it is possible to rewrite the expression for $\Pi_{\mathrm{acc}}$ in the following way:
\begin{equation}
\label{piAccGen}
    \Pi_{\mathrm{acc}}(t)=\frac{4 g}{B(t)} v(t) n(t),
\end{equation}
where $n (t)$ is the physical number density of the monopole pairs,
\begin{equation}
\label{numberdensity}
    n (t) = \frac{1}{a(t)^3} \int_{-\infty}^{t} dt' a(t')^3 \Gamma(t') .
\end{equation}
Hereafter we assume $n \propto a^{-3}$ at times $t > t_i$, namely, that there is no further monopole production after the magnetic fields have switched on. This is a good approximation also for monopoles produced by the magnetic field, since in such a case the monopole population is dominated by those produced at $t \sim t_i$, as we will discuss in the next section.

Using for $\Pi_{\mathrm{red}}$ the definition in Eq.~\eqref{PiRed}, we can express the ratio $\Pi_{\mathrm{acc}} / \Pi_{\mathrm{red}}$ as:
\begin{equation}
\label{piAccRed}
    \frac{\Pi_{\mathrm{acc}}(t)}{\Pi_{\mathrm{red}}(t)} = \frac{g}{B(t) H(t)} v(t) n(t).
\end{equation}
In the case $\Pi_{\mathrm{acc}} / \Pi_{\mathrm{red}} \ll 1$ the solution of Eq.~\eqref{piEq} gives simply $\rho_{B} \propto a^{-4}$, i.e. the energy density of the magnetic field redshifts as radiation.
In the opposite case $\Pi_{\mathrm{acc}}/\Pi_{\mathrm{red}} \gg 1$, the back-reaction on the magnetic fields due to the monopole acceleration is non-negligible and the energy of the fields is transferred to the monopoles at a time scale shorter than the Hubble time. 

While the monopoles interact with the plasma, the energy given to the monopoles by the magnetic fields is further passed on to the plasma.
If the interactions are efficient and in the absence of mechanisms for the regeneration, the magnetic fields would quickly decay away. In this case, the condition $\Pi_{\mathrm{acc}}/\Pi_{\mathrm{red}} \gg 1$, corresponds to a rapid dispersion of the energy of the magnetic fields into the plasma and the disappearing of the fields. 
On the other hand, as discussed in \cite{TPB, Vachaspati}, if the interaction with the plasma is negligible, the kinetic energy of the monopoles are eventually transferred back to the magnetic fields, giving rise to oscillatory behaviors.
Such oscillations can happen at late times, i.e. $T < 1~\mathrm{MeV}$, after the cosmological $e^+ e^-$ annihilation drastically reduces the number density of the scatterers, and possibly at early times when $v \simeq v_{\mathrm{H}}$ and the interaction rate is negligible compared to the Hubble rate.

Without the necessity of specifying the production mechanism for the monopoles, it is then possible to give bounds on their number density from the persistence of the primordial magnetic fields still today. This can be done requiring that the condition $\Pi_{\mathrm{acc}} / \Pi_{\mathrm{red}} \ll 1$ holds when the interaction with the particles of the plasma is non-negligible\footnote{If the initial magnetic field is extremely strong, then the leftover after the damping by the monopoles may serve as the present-day cosmological magnetic fields. However we do not investigate such a possibility in this paper.}.

We show in Figure~\ref{ratio} the evolution of the ratio $\Pi_{\mathrm{acc}} / \Pi_{\mathrm{red}}$ from time $t_i$ for different values of the mass. 
The monopole velocity necessary for computing $\Pi_{\mathrm{acc}} / \Pi_{\mathrm{red}}$ is taken from Figure~\ref{velocity}, which was obtained by solving the equation of motion Eq.~\eqref{eqVelocity}.
We ignore back-reaction on the $B$ field and we apply the relations in Eq.~\eqref{HTevolution} and Eq.~\eqref{aTdom} to rewrite all the quantities that present a time dependence in terms of $H(t)$. 
Since $\Pi_{\mathrm{acc}} / \Pi_{\mathrm{red}} \propto n$, we have normalized the value of $\Pi_{\mathrm{acc}} / \Pi_{\mathrm{red}}$ in the plot by $n_0$ so that its value is independent of the monopole number density today.
The parameter choice and the mass for each colored curve are the same as for Figure \ref{velocity}. Notice that the blue and purple curves overlap with each other almost everywhere in the plot.

For $t>t_{\mathrm{dom}}$ the monopoles move at the terminal velocity set by the interactions with the plasma shown in Eq.~\eqref{vTermPlasma}. In this case the expression for the ratio is:
\begin{equation}
\label{ratioDuringPlasma}
    \frac{\Pi_{\mathrm{acc}}}{\Pi_{\mathrm{red}}} \simeq \frac{g^2}{f_{\mathrm{p}} H} n .
\end{equation}
During radiation domination, the ratio constantly grows as $\Pi_{\mathrm{acc}}/\Pi_{\mathrm{red}} \propto a \propto H^{-1/2}$. Such behavior is shown in the right part of Figure~\ref{ratio} as the grey line and it is independent of the monopole mass.

For $t<t_{\mathrm{dom}}$, the time evolution can be more complicated and exhibits mainly three kinds of behaviors. The first case is realized for monopoles light enough to be initially accelerated to relativistic velocities, i.e. $v \simeq 1$. In this case the Hubble friction dominates over the friction from the plasma. As long as the monopoles maintain a relativistic velocity, the expression for the ratio is:
\begin{equation}
\label{case1}
     \frac{\Pi_{\mathrm{acc}}}{\Pi_{\mathrm{red}}} \simeq \frac{g}{B H} n .
\end{equation}
This ratio scales as $\Pi_{\mathrm{acc}}/\Pi_{\mathrm{red}} \propto a^{1/2} \propto H^{-1/3}$ and increases with time. Eq.~\eqref{case1} corresponds to the growing segments of the purple, blue, green, and orange curves in the left part of Figure~\ref{ratio}. For the parameter choice in the plot, the velocity is relativistic soon after $t_i$ for $m \lesssim 10^{16}~\mathrm{GeV}$.

The second case is when the Hubble friction is dominant over the plasma friction and the monopoles present non-relativistic velocities. In this case, the velocity is given by $v \sim gB / \left( mH \right)$ and the ratio is constant in time:
\begin{equation}
\label{case2}
     \frac{\Pi_{\mathrm{acc}}}{\Pi_{\mathrm{red}}} \simeq \frac{g^2}{m H^2} n .
\end{equation}
This case corresponds to the horizontal segments of the brown, red, and orange curves of Figure~\ref{ratio}.

The last case is realized when the monopoles achieve the terminal velocity set by the interaction with the plasma. The expression for the ratio in this case is the same as that in Eq.~\eqref{ratioDuringPlasma}, but during reheating this scales as $\Pi_{\mathrm{acc}}/\Pi_{\mathrm{red}} \propto a^{-3/4} \propto H^{1/2}$, decreasing in time.
This case is shown in Figure~\ref{ratio} as the decreasing segments of the colored curves.

During the reheating epoch, the $\Pi_{\mathrm{acc}}/\Pi_{\mathrm{red}}$ ratio given by monopoles with masses $m < \bar{m}$ is maximized at the time when the monopoles turn non-relativistic.
On the other hand, for $m > \bar{m}$, the ratio $\Pi_{\mathrm{acc}}/\Pi_{\mathrm{red}}$ exhibits a plateau-like behavior while the monopole velocity follows $v_H$ and is non-relativistic.

In correspondence to the jump from a relativistic velocity to a non-relativistic velocity shown in Figure~\ref{velocity} for $m < \bar{m}$, a jump in the value of the ratio is seen in Figure~\ref{ratio} for the purple and blue curves.
Such a jump corresponds to a sudden transition during reheating from the value of the ratio shown in Eq.~\eqref{case1} to that of Eq.~\eqref{ratioDuringPlasma}.

Below we obtain bounds on the cosmic abundance of monopoles by requiring that $\Pi_{\mathrm{acc}}/\Pi_{\mathrm{red}}$ stays smaller than unity during radiation domination and reheating.

\subsection{During radiation domination}
\label{during3}

The analysis that we present for the bound during radiation domination follows the work in \cite{Vachaspati}.
For times $t>t_{\mathrm{dom}}$, we express the value of $\Pi_{\mathrm{acc}}/\Pi_{\mathrm{red}}$ through the result shown in Eq.~\eqref{ratioDuringPlasma}. Using the expression for $f_{\mathrm{p}}$ given in Eq.~\eqref{effep}, we can rewrite the ratio $\Pi_{\mathrm{acc}}/\Pi_{\mathrm{red}}$ as:
\begin{equation}
    \frac{\Pi_{\mathrm{acc}}}{\Pi_{\mathrm{red}}} \simeq \frac{16 \pi^2}{e^2 \mathcal{N}_c T^2 H} n .
\end{equation}
Using $H \simeq (\pi / \sqrt{90}) g_*^{1/2} T^2 / M_{\mathrm{Pl}}$, $n \propto a^{-3}$ and the relation between the scale factor and the temperature in Eq.~\eqref{gstar1}, the expression for $\Pi_{\mathrm{acc}}/\Pi_{\mathrm{red}}$ becomes:
\begin{equation}
\label{pipiDuringGen}
    \frac{\Pi_{\mathrm{acc}}}{\Pi_{\mathrm{red}}} \simeq \frac{48 \sqrt{10} \pi g_{*s} M_{\mathrm{Pl}}}{e^2 \mathcal{N}_c g_*^{1/2}  g_{* s, 0}} \frac{n_0}{T_0^3} \frac{1}{T} ,
\end{equation}
where $g_{* s, 0} \simeq 3.9$. As shown in Figure~\ref{ratio}, the ratio increases with time during radiation domination.
The expression in Eq.~\eqref{effep} for the friction assumes relativistic plasma particles, and hence our analysis is valid up to the time of $e^+ e^-$ annihilation, namely, when $T \sim 1~\mathrm{MeV}$. After the annihilation, the number of charged particles in the plasma decreases by a factor $10^{-10}$ \cite{TDGK} and thus the monopoles cannot give away their energy effectively to the plasma.
For this value of the temperature we have $\mathcal{N}_c \sim g_* \simeq g_{*s} \simeq 10.75$. Therefore, the maximum value of the ratio during radiation domination is:
\begin{equation}
\label{pipi1mev}
    \frac{\Pi_{\mathrm{acc}}}{\Pi_{\mathrm{red}}} (T = 1~\mathrm{MeV}) \simeq \frac{n_0}{10^{-21}~\mathrm{cm}^{-3}} .
\end{equation}
The survival of the primordial magnetic field then requires $\Pi_{\mathrm{acc}} / \Pi_{\mathrm{red}} (T = 1~\mathrm{MeV}) \lesssim 1$. This yields the Primordial Parker Bound of \cite{Vachaspati}:
\begin{equation}
    n_0 \lesssim 10^{-21}~\mathrm{cm}^{-3} .
\end{equation}
The bound does not depend on $B_0$, $g$, or $m$. It should also be noted that this is a bound on the average monopole number density in the universe.

Introducing the flux of monopoles at time $t$ as $F(t) = n (t) v(t) / 4 \pi$, we can express the above bound in terms of the present-day monopole flux $F_0$:
\begin{equation}
\label{PrimordialParker}
    F_0 \lesssim 10^{-15}~\mathrm{cm}^{-2} \mathrm{sr}^{-1} \mathrm{s}^{-1}  \left( \frac{v_0}{10^{-3}} \right) .
\end{equation}
Here, we use the virial velocity in the Galaxy $10^{-3}$ as a reference value \cite{KolbTurner}, although as we mentioned above this result applies to the average monopole flux in the universe.
Let us also notice that we define the flux as that of only monopoles (or antimonopoles) and thus our results differ by a factor $2$ from those for the flux of both monopoles and antimonopoles. However, such a difference is negligible for the order-of-magnitude bounds we derive in this work.

Strictly speaking, the above analysis is valid only for monopoles with sub-Planckian masses, cf. Eq.~\eqref{subplanck}. For monopoles that present masses $m > M_{\mathrm{Pl}} \mathcal{N}_c / g_{*}^{1/2}$, the Hubble friction term of the equation of motion is dominant over the drag force of the plasma even during radiation domination. Such monopoles never achieve the terminal velocity set by the plasma, and hence the bound in Eq.~\eqref{PrimordialParker} must be modified.

\subsection{Before radiation domination}
\label{before3}

Bounds on the monopole number density can also be derived based on the survival of primordial magnetic fields during the reheating epoch.
Here we assume the cosmological plasma during reheating to be in thermal equilibrium, however let us remark that this assumption leads to a conservative bound on the monopole abundance.
Without the assumption of thermal equilibrium, the number density of the particles of the plasma is not related to the mean energy of the particles.
Inflaton decay results in an initially dilute plasma that contains a small number of very energetic particles that are not in thermal equilibrium \cite{reheating}. Under these conditions the monopoles are more easily accelerated by the magnetic fields than in the case of thermal equilibrium.\footnote{The rough estimate of the drag force is $f_{\mathrm{p}} \sim n \sigma \Delta p$, where $\sigma \sim e^2 g^2 E^{-2}$ is the cross section of the interaction with the particles of the plasma with mean energy $E$ and $\Delta p \sim E$ is the exchanged momentum. Since the number density $n$ is smaller and $\sigma \Delta p \propto E^{-1}$, the drag force is also smaller in the absence of thermal equilibrium.}
Thus, the drag force is smaller and the monopole velocity larger. 
Hence, the damping rate of the magnetic field, $\Pi_{\mathrm{acc}} \propto v$, turns out to be larger than in the case of thermal equilibrium, and the resulting bound on the monopole abundance can become stronger. We leave the case of a non-thermal plasma for future analysis.

As discussed below Eq.~\eqref{piAccRed}, the survival of the primordial magnetic fields requires the condition $\Pi_{\mathrm{acc}}/\Pi_{\mathrm{red}} \ll 1$ to be satisfied while the monopoles frequently interact with the plasma particles. Hence we can restrict ourselves to times when the monopole velocity is controlled by the plasma friction, instead of the Hubble friction\footnote{If $\Pi_{\mathrm{acc}}/\Pi_{\mathrm{red}} > 1$ during $t < t_*$, the energy rapidly oscillates between the magnetic field and monopoles, whose effect is to modify the redshifting of the magnetic energy density from the usual $\rho_{\mathrm{B}} \propto a^{-4}$ \cite{Vachaspati}.}.
And since we are interested in times after the magnetic fields have been generated, we focus on the regime $ \max \left \{ t_*, t_i \right \} \leq t \leq t_{\mathrm{dom}}$, where $t_*$ is defined in Eq.~\eqref{fstarcondition}. 
The expression for $\Pi_{\mathrm{acc}}/\Pi_{\mathrm{red}}$ during this regime is given in Eq.~\eqref{ratioDuringPlasma}, which decreases with the Hubble scale as $\propto H^{1/2}$. Therefore, in order to derive the strongest bound from the reheating epoch, we should evaluate $\Pi_{\mathrm{acc}}/\Pi_{\mathrm{red}}$ at $t = \max \left \{ t_*, t_i \right \}$.\footnote{For $m < \bar{m}$, the ratio $\Pi_{\mathrm{acc}} / \Pi_{\mathrm{red}}$ actually continues to increase after $t = t_*$ for about a Hubble time. However, we use simply Eq.~\eqref{ratioDuringPlasma} to derive our bound because the value of the ratio does not change substantially within a Hubble time. This leads to a conservative bound on the monopole abundance.}

Considering that $n \propto a^{-3}$, the expression for the value of the ratio at $t = \max \left \{ t_*, t_i \right \}$ is:
\begin{equation}
\label{HiEq}
    \frac{\Pi_{\mathrm{acc}}}{\Pi_{\mathrm{red}}} (t = \max \left \{ t_*, t_i \right \} ) \simeq \frac{16 \pi^2}{e^2 \mathcal{N}_{c, \mathrm{dom}} H_{\mathrm{dom}}^{3/2} T_{\mathrm{dom}}^{2}} \left( \frac{a_0}{a_{\mathrm{dom}}} \right)^3 \left( \min \left \{ H_*, H_i \right \} \right)^{1/2} n_0 .
\end{equation}
Now we derive the bounds on the monopole abundance in both the cases $H_i < H_*$ and $H_i > H_*$. Once the value of $H_*$ is fixed, the bound in the case $H_i < H_*$ is always weaker than the bound for $H_i > H_*$.

\subsubsection[Case with $H_i < H_*$]{Case with $\mathbf{H_i < H_*}$}

For $H_i < H_*$, we rewrite the condition $\Pi_{\mathrm{acc}} / \Pi_{\mathrm{red}} \lesssim 1$ on Eq.~\eqref{HiEq} as a bound on the average monopole number density in the present universe using the relations in Eq.~\eqref{aTdom}:
\begin{equation}
    n_0 \lesssim 10^{-16}~\mathrm{cm}^{-3} \left( \frac{\mathcal{N}_{c, \mathrm{dom}}}{100} \right)
    \left( \frac{T_{\mathrm{dom}}}{10^{6}~\mathrm{GeV}} \right)^2 \left( \frac{10^{4}~\mathrm{GeV}}{H_i} \right)^{1/2} .
\end{equation} 
As a reference value for $H_i$, here we chose $10^{4}~\mathrm{GeV}$ which is the reference value for $H_*$ in Eq.~\eqref{HstarBoth}.
We express the result also in terms of the monopole flux today:
\begin{equation}
    F_0 \lesssim 10^{-10}~\mathrm{cm}^{-2} \mathrm{sr}^{-1} \mathrm{s}^{-1}  \left( \frac{v_0}{10^{-3}} \right) \left( \frac{\mathcal{N}_{c, \mathrm{dom}}}{100} \right)
    \left( \frac{T_{\mathrm{dom}}}{10^{6}~\mathrm{GeV}} \right)^2 \left( \frac{10^{4}~\mathrm{GeV}}{H_i} \right)^{1/2} .
\end{equation}

\subsubsection[Case with $H_i > H_*$]{Case with $\mathbf{H_i > H_*}$}

We get a bound for the average monopole number density in the present universe applying the condition $\Pi_{\mathrm{acc}} / \Pi_{\mathrm{red}} \lesssim 1$ on Eq.~\eqref{HiEq} to the case $H_i > H_*$ and using the expression for $H_*$ shown in Eq.~\eqref{HstarBoth}:
\begin{equation}
    n_0 \lesssim 
    \begin{dcases}
    10^{-16}~\mathrm{cm}^{-3}  \left( \frac{B_0}{10^{-15}~\mathrm{G}} \right)^{3/5} \left( \frac{T_{\mathrm{dom}}}{10^{6}~\mathrm{GeV}} \right) \left( \frac{10}{g} \right)^{3/5} &,~~ m \ll \bar{m} , \\
    10^{-16}~\mathrm{cm}^{-3}  \left( \frac{m}{10^{14}~\mathrm{GeV}} \right) \left( \frac{T_{\mathrm{dom}}}{10^{6}~\mathrm{GeV}} \right) \left( \frac{10}{g} \right)^{2} &,~~ m \gg \bar{m} .
    \end{dcases}
\end{equation}
Here we drop the dependence on $\mathcal{N}_{c,\mathrm{dom}}$, because the final results depend weakly on its value. The bound for the monopole flux today is:
\begin{equation}
\label{perri}
    F_0 \lesssim
    \begin{dcases}
    10^{-10}~\mathrm{cm}^{-2}\mathrm{sr}^{-1} \mathrm{s}^{-1}  \left( \frac{v_0}{10^{-3}} \right) \left( \frac{B_0}{10^{-15}~\mathrm{G}} \right)^{3/5} \left( \frac{T_{\mathrm{dom}}}{10^{6}~\mathrm{GeV}} \right) \left( \frac{10}{g} \right)^{3/5} &,~~ m \ll \bar{m} , \\
    10^{-10}~\mathrm{cm}^{-2} \mathrm{sr}^{-1} \mathrm{s}^{-1}  \left( \frac{v_0}{10^{-3}} \right) \left( \frac{m}{10^{14}~\mathrm{GeV}} \right) \left( \frac{T_{\mathrm{dom}}}{10^{6}~\mathrm{GeV}} \right) \left( \frac{10}{g} \right)^{2} &,~~ m \gg \bar{m} .
    \end{dcases}
\end{equation}

Recall that $\bar{m}$ divides monopoles that are relativistic ($m \ll \bar{m}$) or non-relativistic ($m \gg \bar{m}$) when the interactions with the plasma become important.
For $m \ll \bar{m}$ the bound depends on the value of the magnetic field today, while it is independent of the monopole mass. 
On the contrary, for $m \gg \bar{m}$ the result is independent of $B_0$, while it is proportional to the mass of the monopoles.

\vspace{5pt}

\begin{figure}[!t]
  \centering
  \includegraphics[width=\textwidth]{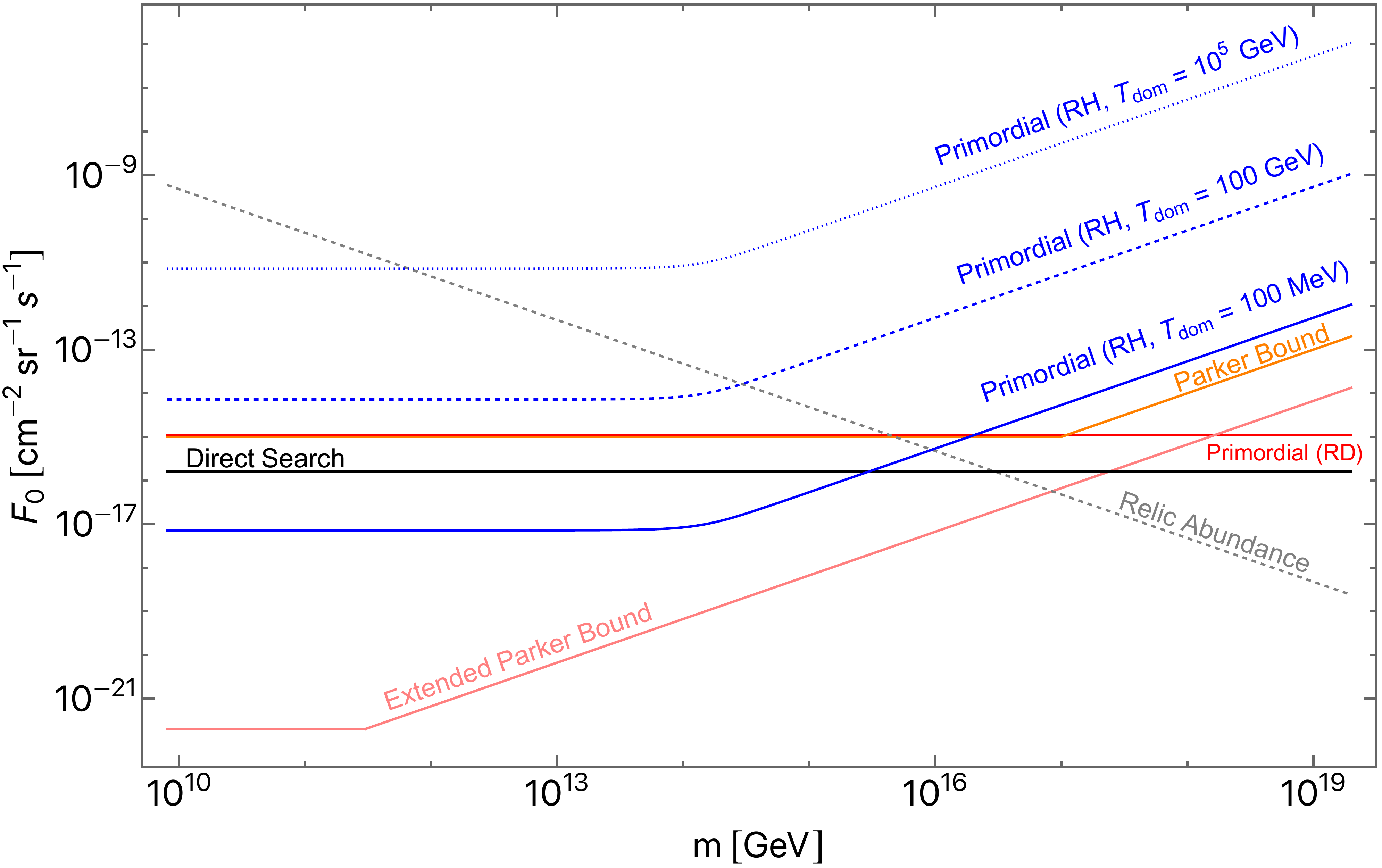}
\caption{Upper bounds on the magnetic monopole flux today. Here  $g = 2 \pi /e$, $B_0 = 10^{-15}~\mathrm{G}$ and $v_0 = 10^{-3}$. Blue: bounds from primordial magnetic fields during the reheating epoch shown in Eq.~\eqref{perri}, for reheating temperatures $T_{\mathrm{dom}} = 100~\mathrm{MeV}$ (solid curve), $100~\mathrm{GeV}$ (dashed curve), $10^5~\mathrm{GeV}$ (dotted curve).
Red: the bound from primordial magnetic fields during radiation domination shown in Eq.~\eqref{PrimordialParker}. Orange: the original Parker bound. Pink: the extended Parker bound. Black: direct search limit from the MACRO experiment. Dashed grey: the cosmological abundance bound.}
\label{Perri}
\end{figure}

In Figure \ref{Perri} we compare various upper bounds on the monopole flux today as functions of the monopole mass. The blue curves show the bounds that we derived in Eq.~\eqref{perri} from the survival of the primordial magnetic fields during reheating for three different values of $T_{\mathrm{dom}} = 100~\mathrm{MeV}$ (solid curve), $100~\mathrm{GeV}$ (dashed curve), $10^5~\mathrm{GeV}$ (dotted curve).
The red line shows the bound in Eq.~\eqref{PrimordialParker} from the survival of primordial magnetic fields during radiation domination, which was first obtained in \cite{Vachaspati}. The orange line corresponds to the original Parker bound from the survival of the Galactic magnetic field \cite{TPB}. The pink line corresponds to the ``extended Parker bound'' that has been derived from the survival of the Galactic seed field by \cite{AFFTWT}.\footnote{The original and extended Parker bounds in the large $m$ region (where the bounds grow with $m$) are independent of the magnetic field strength, and thus should be equivalent if the other parameters are the same. The lines in the plot do not coincide because they exhibit the results presented in \cite{TPB} and \cite{AFFTWT}, which use slightly different values for the parameters, as well as different rounding methods.}
The black line shows the limit obtained by the MACRO experiment \cite{MACRO}, which corresponds to the strongest bound from the direct search of non-relativistic monopoles. The dashed grey line shows the cosmological abundance bound from the requirement that the monopole energy density is smaller than the total energy density of the universe \cite{Preskill, KolbTurner}.
In the plot we assume $g = 2 \pi /e$, $B_0 = 10^{-15}~\mathrm{G}$ and the reference value of the monopole velocity today $v_0 = 10^{-3}$.

In the plot we have displayed the various bounds for comparison purpose, however we should remark that their targets are different: the bounds based on primordial magnetic fields and the relic abundance constrain the average monopole number density in the universe, while the bounds from Galactic fields (the original and extended Parker bounds) and direct searches constrain the monopole density inside the Galaxy. If the monopoles are clustered with the Galaxy, their local density in the Galaxy can be significantly larger than the average density in the universe; in such a case the bounds on the local density translate into much stronger bounds on the average density.

As shown in the plot, for a sufficiently small $T_{\mathrm{dom}}$, our bound in Eq.~\eqref{perri} from the analysis during reheating becomes stronger than the original Parker bound and the limits from direct searches. For a GUT scale monopole, our bound is comparable to the original Parker bound for $T_{\mathrm{dom}} \sim 1~\mathrm{GeV}$.
The bound during reheating can also be stronger than that during radiation domination, Eq.~\eqref{PrimordialParker}, for $T_{\mathrm{dom}} \lesssim 10~\mathrm{GeV}$ and in the low-mass range.

\section{Monopoles produced by primordial magnetic fields}
\label{prodField}

In the previous section, we derived bounds on the abundance of monopoles without specifying their origin. Here we focus on monopoles that are Schwinger-produced by the primordial magnetic field itself, and study whether the magnetic field is dissipated by the monopole acceleration.
Here we note that, even in the absence of any initial monopole population, the strong magnetic fields in the early universe can trigger the monopole pair production. By studying this process, we derive further constraints on monopoles, which also serve as the most conservative condition for the survival of primordial magnetic fields.

The Schwinger effect describes the production of particle-antiparticle pairs in an external field. The analysis by Schwinger \cite{Schwinger} assumed weak couplings, however this is not the case for magnetic monopoles which, due to the Dirac quantization condition, have strong magnetic couplings.
The rate of monopole-antimonopole pair production at arbitrary coupling in a static magnetic field has been derived in \cite{AM, AAM} through an instanton method:
\begin{equation}
\label{rate}
    \Gamma = \frac{(g B)^{2}}{(2 \pi)^{3}} \exp \left[-\frac{\pi m^{2}}{g B}+\frac{g^{2}}{4}\right] .
\end{equation}
This result is valid under the following weak field conditions:
\begin{subequations} 
\label{InstantonLimitBoth}
    \begin{equation}
        B \lesssim \frac{m^{2}}{g} ,
    \end{equation}
    \begin{equation}
    \label{InstantonLimit}
        B \lesssim \frac{4\pi m^{2}}{g^{3}} .
    \end{equation}
\end{subequations}
The second condition is stricter than the first one if $g \gg 1$, and suggests that the instanton computation is valid when the exponent of the expression in Eq.~\eqref{rate} is negative. 

After the process of magnetogenesis, the primordial magnetic fields redshift as $B \propto a^{-2}$. Thus, it suffices to assume that the weak field condition Eq.~\eqref{InstantonLimit} is verified at time $t_i$.
Defining:
\begin{equation}
    \mathscr{I} = \frac{4 \pi m^{2}}{g^{3} B_i} ,
\end{equation}
we rewrite the weak field condition Eq.~\eqref{InstantonLimit} as:
\begin{equation}
\label{InstantonFinal}
   \mathscr{I} \gtrsim 1 .
\end{equation}

Using Eqs.~\eqref{aTdom} and~\eqref{aidom}, the initial amplitude of the primordial magnetic field can be written in terms of the field strength today as:
\begin{equation}
\label{Bi}
    B_i = B_0 \left(\frac{a_0}{a_i}\right)^2 \simeq 10^{43}~\mathrm{G} \left( \frac{B_0}{10^{-15}~\mathrm{G}} \right) \left( \frac{H_i}{10^{14}~\mathrm{GeV}} \right) \max \left \{ \left( \frac{H_i}{H_{\mathrm{dom}}} \right)^{1/3}\ ,\ 1 \right \} .
\end{equation}
Here, the first expression within the curly brackets corresponds to the case with $t_i < t_{\mathrm{dom}}$ and the second to $t_i > t_{\mathrm{dom}}$.
Substituting the expression for $B_i$ into the definition of $\mathscr{I}$, we get:
\begin{equation}
    \mathscr{I} \simeq \left( \frac{10}{g} \right)^3 \left( \frac{10^{-15}~\mathrm{G}}{B_0} \right) \left( \frac{10^{14}~\mathrm{GeV}}{H_i} \right) \left( \frac{m}{10^{12}~\mathrm{GeV}} \right)^{2} \min \left \{ \left( \frac{H_{\mathrm{dom}}}{H_i} \right)^{1/3}\ ,\ 1 \right \} ,
\end{equation}
and we can express the weak field condition in Eq.~\eqref{InstantonLimit} as a lower limit on the mass of the monopoles in terms of the Hubble rate at magnetogenesis:
\begin{equation}
\label{weakfieldMass}
    m \gtrsim 10^{12}~\mathrm{GeV} \left( \frac{g}{10} \right)^{3/2} \left( \frac{B_0}{10^{-15}~\mathrm{G}} \right)^{1/2} \left( \frac{H_i}{10^{14}~\mathrm{GeV}} \right)^{1/2} \max \left \{ \left( \frac{H_i}{H_{\mathrm{dom}}} \right)^{1/6}\ ,\ 1 \right \} .
\end{equation}

The number density of monopoles pair-produced by the magnetic fields can be obtained using the expression in Eq.~\eqref{numberdensity} and substituting the production rate shown in Eq.~\eqref{rate}.
Because the production rate presents an exponential dependence on the magnetic fields, the monopoles are produced predominantly within an interval $\Delta t_{\Gamma_i} \sim |\Gamma_i / \dot{\Gamma}_i | \simeq (g B_i)/(2 \pi m^2 H_i)$ after $t_i$ \cite{PrimMag}. Therefore, we can approximately express the number density at times around $t_i$ as:
\begin{equation}
    n_i \simeq \frac{g \Gamma_i B_i}{2 \pi m^2 H_i} .
\end{equation}
Using $n \propto a^{-3}$, the monopole number density today is thus written as:
\begin{equation}
\label{n0ni}
    n_0 \simeq \left( \frac{a_i}{a_0} \right)^3 \frac{g \Gamma_i B_i}{2 \pi m^2 H_i} .
\end{equation}

It was pointed out in \cite{PrimMag} that Eq.~\eqref{InstantonLimit} also gives an absolute upper bound on the initial amplitude of primordial magnetic fields: saturating this bound leads to either an overproduction of monopoles in the universe, or a self-screening of the magnetic field. Below we revisit the magnetic field bound in light of the constraints derived in the previous section.
We show that for the monopoles produced by the primordial magnetic fields, the bounds on the monopole flux approximately reduce to the weak field condition in Eq.~\eqref{InstantonLimit}.

\subsection{During radiation domination}

We now substitute the expression for the number density of monopoles produced by the Schwinger effect in Eq.~\eqref{n0ni} into the expression for the maximum of $\Pi_{\mathrm{acc}}/\Pi_{\mathrm{red}}$ during radiation domination shown in Eq.~\eqref{pipi1mev}. Making use of the relations in Eqs.~\eqref{Bi}, \eqref{aTdom},~\eqref{aidom}, we can rewrite the expression for the ratio as:
\begin{equation}
\begin{split}
    \frac{\Pi_{\mathrm{acc}}}{\Pi_{\mathrm{red}}} (T = 1~\mathrm{MeV}) \simeq \Tilde{x}_D \exp{ \left( - \frac{g^2}{4} \left( \mathscr{I} -1 \right) \right)} ,
\end{split}
\end{equation}
where we define:
\begin{equation}
    \Tilde{x}_D = \left( \frac{g}{10} \right)^{3} \left( \frac{B_0}{10^{-15}~\mathrm{G}} \right)^3 \left( \frac{10^{16}~\mathrm{GeV}}{m} \right)^2 \left( \frac{H_i}{10^{14}~\mathrm{GeV}} \right)^{1/2} \max \left \{ \left( \frac{H_i}{H_{\mathrm{dom}}} \right)^{1/2}\ ,\ 1 \right \} .
\end{equation}
Under the requirement of negligible back-reaction on the primordial magnetic fields, \\ $\Pi_{\mathrm{acc}} / \Pi_{\mathrm{red}} (T = 1~\mathrm{MeV}) \lesssim 1$, we get:
\begin{equation}
\label{boundDuring}
    \Tilde{x}_D \exp{ \left( - \frac{g^2}{4} \left( \mathscr{I} -1 \right) \right)} \lesssim 1 .
\end{equation}

\begin{figure}[!t]
  \centering
  \includegraphics[width=0.7\textwidth]{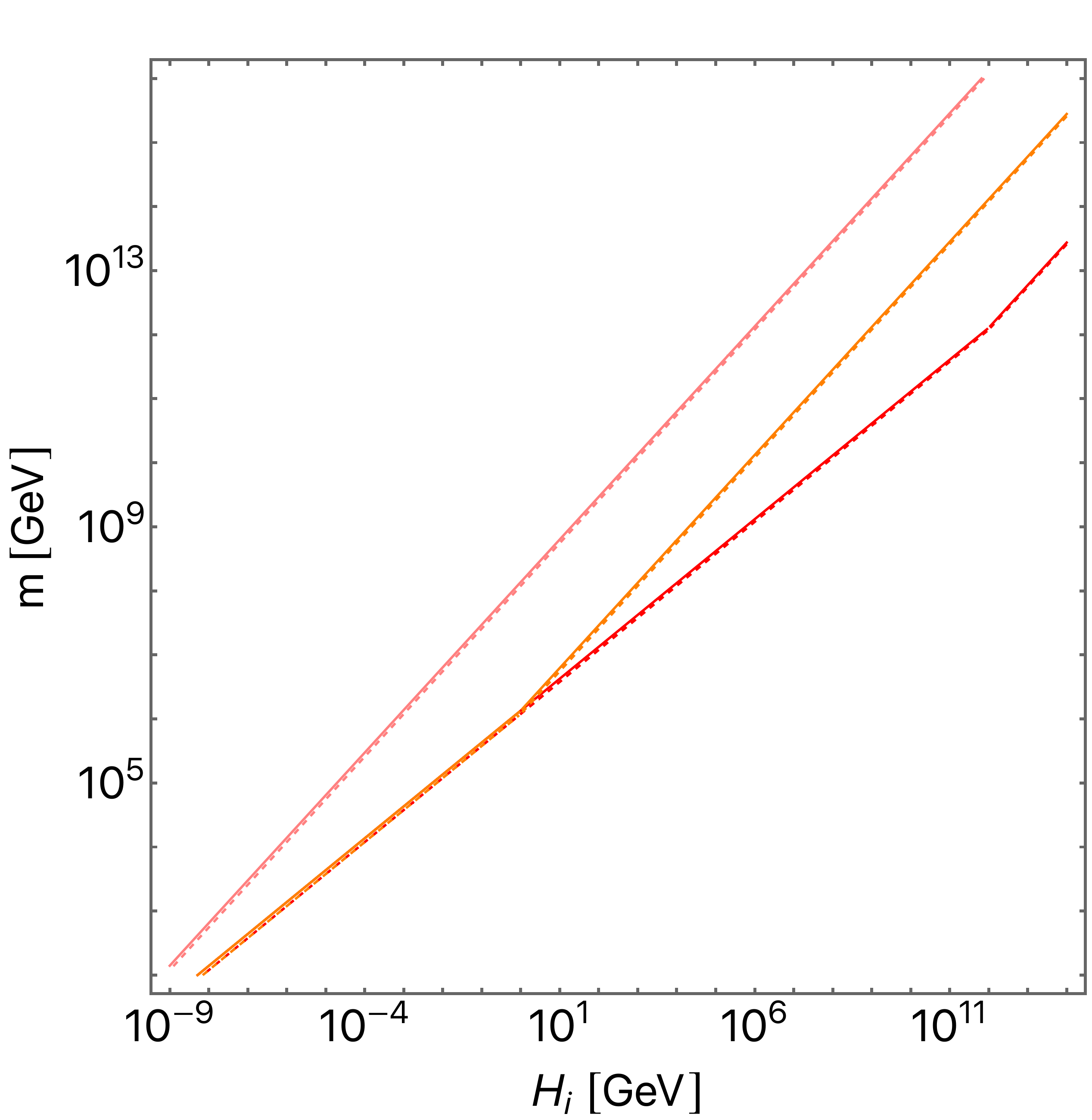}
\caption{Lower bounds for the monopole mass as a function of the Hubble scale at magnetic field generation $H_i$. The plot shows the comparison between the weak field condition in~\eqref{weakfieldMass} (dashed curves) and the bound from requiring negligible back-reaction on the primordial magnetic fields during radiation domination, shown in~\eqref{boundDuring} (solid curves). Here $g = 2 \pi /e$ and $B_0 = 10^{-15}~\mathrm{G}$. The results are shown for three different values of the reheating temperature $T_{\mathrm{dom}}$; from bottom to top, red: $10^{15}~\mathrm{GeV}$, orange: $10^{9}~\mathrm{GeV}$, pink: $10^{3}~\mathrm{GeV}$. For $H_i < H_{\mathrm{dom}}$, the results are independent from $T_{\mathrm{dom}}$.}
\label{During}
\end{figure}

In Figure~\ref{During} we show in solid curves the bound in Eq.~\eqref{boundDuring}. We also plot the weak field condition in Eq.~\eqref{weakfieldMass} in dashed curves. The region below the curves is not compatible with the survival of the primordial magnetic fields, namely, the curves give lower bounds on the monopole mass. For the computation we assume $g = 2 \pi /e$ and $B_0 = 10^{-15}~\mathrm{G}$.
The bound is shown for different values of $T_{\mathrm{dom}}$ (from bottom to top, red: $10^{15}~\mathrm{GeV}$, which corresponds to $H_{\mathrm{dom}} = 10^{12}~\mathrm{GeV}$; orange: $10^{9}~\mathrm{GeV}$, which corresponds to $H_{\mathrm{dom}} = 1~\mathrm{GeV}$; pink: $10^{3}~\mathrm{GeV}$, which corresponds to $H_{\mathrm{dom}} = 10^{-12}~\mathrm{GeV}$). 
In cases where the magnetic fields are produced after reheating, i.e. $t_i > t_{\mathrm{dom}}$, the bound is independent of $T_{\mathrm{dom}}$; this is seen in the plot as the red and orange curves overlapping on the low-$H_i$ end.

Due to the exponential dependence on $\mathscr{I}$ in Eq.~\eqref{boundDuring}, this condition reduces approximately to the weak field condition Eq.~\eqref{weakfieldMass}, as seen in the plot. This can be easily shown by computing the natural logarithm of the expression and rewriting it as:
\begin{equation}
    \mathscr{I} \gtrsim 1 + \frac{4}{g^2} \ln{\Tilde{x}_D} .
\end{equation}

The expression now takes a form similar to that for the weak field condition in Eq.~\eqref{InstantonFinal} with an additional logarithmic factor. For values of $H_i$ that saturates the upper bound on the inflation scale $H \lesssim 10^{14}~\mathrm{GeV}$ \cite{Planck} and assuming for $B_0$ the reference value of $10^{-15}~\mathrm{G}$, the logarithmic term can become comparable to $1$ only for very low monopole masses and very small values of $T_{\mathrm{dom}}$. In any case, the corrections are generically less than order unity and negligible for an estimate at the level of the order of magnitude.
In summary, the bounds on the monopole abundance obtained for $t > t_{\mathrm{dom}}$ approximately reduce to the weak field condition in Eq.~\eqref{weakfieldMass}.

\subsection{Before radiation domination}

We now study the back-reaction of pair-produced monopoles on the primordial magnetic field during the reheating epoch. Hence in this subsection we only focus on cases where the magnetic field generation takes place prior to radiation domination.

As explained in Section~\ref{before3}, the monopole bound is obtained by evaluating the ratio $\Pi_{\mathrm{acc}}/\Pi_{\mathrm{red}}$ at the time $t_i$ or $t_*$, whichever is later (see Eq.~\eqref{HstarBoth} for the definition of $H_*$).
We compute the ratio by substituting Eq.~\eqref{n0ni} into Eq.~\eqref{HiEq}.
Taking into account the relations in Eqs.~\eqref{Bi}, \eqref{aTdom},~\eqref{aidom}, we obtain:
\begin{equation}
\begin{split}
\label{notYetT}
    \frac{\Pi_{\mathrm{acc}}}{\Pi_{\mathrm{red}}} \left( t = \max \left \{ t_i, t_* \right \} \right) \simeq\ \Tilde{x}_B \exp{ \left( - \frac{g^2}{4} \left( \mathscr{I} -1 \right) \right)} ,
\end{split}
\end{equation}
where we define:
\begin{equation}
\begin{split}
    \Tilde{x}_B = \left( \frac{g}{10} \right)^3 & \left( \frac{100}{\mathcal{N}_{c, \mathrm{dom}}} \right) \left( \frac{B_0}{10^{-15}~\mathrm{G}} \right)^3 
    \left( \frac{10^{11}~\mathrm{GeV}}{T_{\mathrm{dom}}} \right)^3 \left( \frac{10^{13}~\mathrm{GeV}}{m} \right)^2 \\ & \cdot \left( \frac{H_i}{10^{14}~\mathrm{GeV}} \right)^{3/2}  \min \left \{ \left( \frac{H_*}{H_i} \right)^{1/2} , 1 \right \} .
\end{split}
\end{equation}
We remind the reader that the form of $H_*$ in Eq.~\eqref{HstarBoth} depends on whether the mass of the monopoles is smaller or larger than $\bar{m}$.

Negligible back-reaction on the magnetic fields implies $\Pi_{\mathrm{acc}}/\Pi_{\mathrm{red}} \left( t = \max \left \{ t_i, t_* \right \} \right) \lesssim 1$. From this condition, we get a bound in the $(H_i, m)$ plane similar to that obtained during radiation domination:
\begin{equation}
\label{boundBefore}
    \Tilde{x}_B \exp{ \left( - \frac{g^2}{4} \left( \mathscr{I} -1 \right) \right)} \lesssim 1 .
\end{equation}

\begin{figure}[!t]
  \centering
  \includegraphics[width=0.7\textwidth]{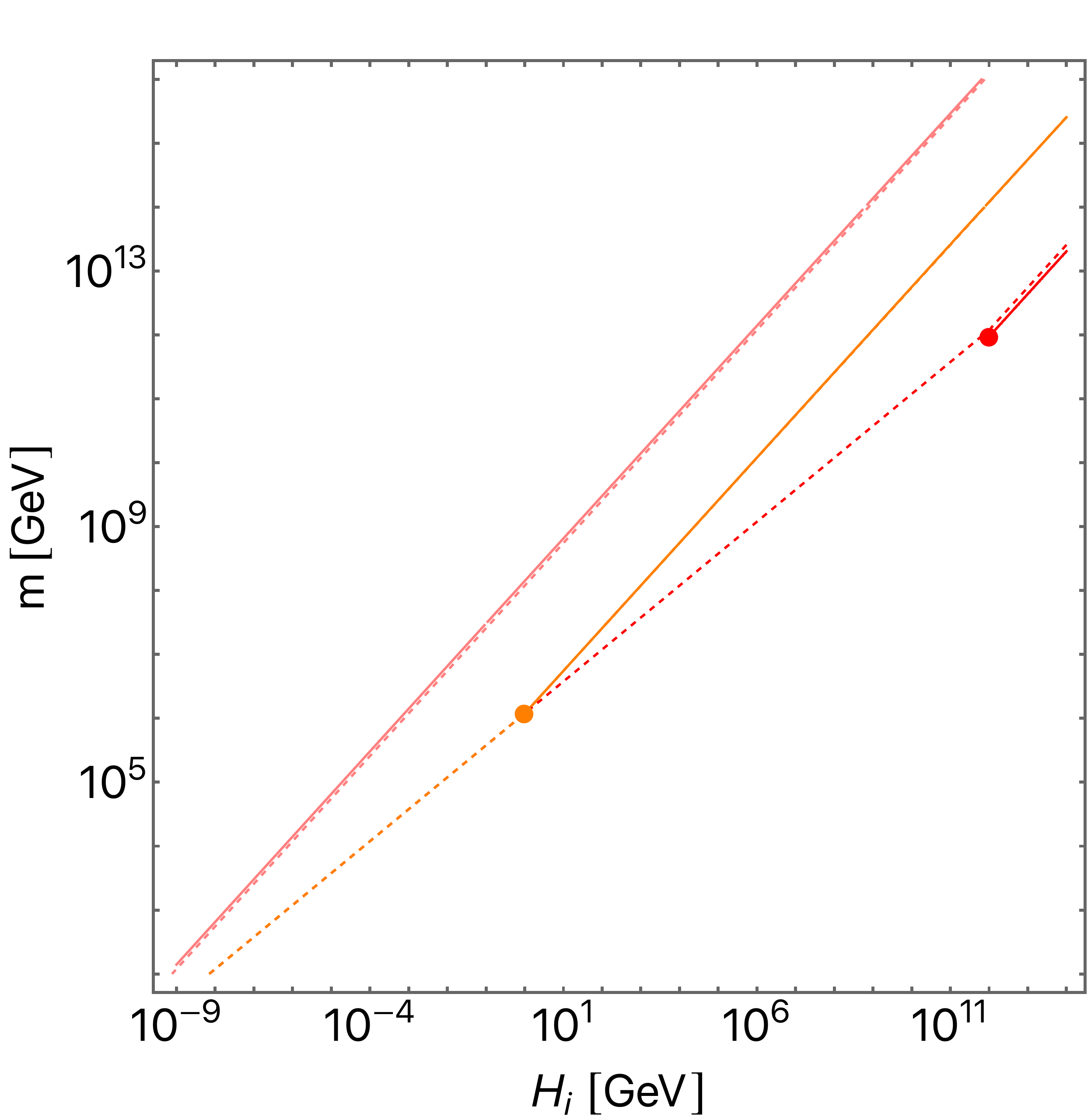}
\caption{Lower bounds for the monopole mass as a function of the Hubble scale at magnetic field generation $H_i$. The plot shows the comparison between the weak field condition in~\eqref{weakfieldMass} (dashed curves) and the bound from requiring negligible back-reaction on the primordial magnetic fields during reheating, shown in~\eqref{boundBefore} (solid curves). Here $g = 2 \pi /e$, $\mathcal{N}_{c, \mathrm{dom}} = 100$ and $B_0 = 10^{-15}~\mathrm{G}$. The results are shown for three different values of the reheating temperature $T_{\mathrm{dom}}$; from bottom to top, red: $10^{15}~\mathrm{GeV}$, orange: $10^{9}~\mathrm{GeV}$, pink: $10^{3}~\mathrm{GeV}$.
The bound of Eq.~\eqref{boundBefore} applies only for $H_i \geq H_{\mathrm{dom}}$. The endpoints of the solid curves correspond to where $H_i = H_{\mathrm{dom}}$.}
\label{Before}
\end{figure}

In Figure~\ref{Before} we show in solid curves the condition in Eq.~\eqref{boundBefore} and in dashed curves the weak field condition given in Eq.~\eqref{weakfieldMass} for different values of $T_{\mathrm{dom}}$ (from bottom to top, red: $10^{15}~\mathrm{GeV}$, which corresponds to $H_{\mathrm{dom}} = 10^{12}~\mathrm{GeV}$; orange: $10^{9}~\mathrm{GeV}$, which corresponds to $H_{\mathrm{dom}} = 1~\mathrm{GeV}$; pink: $10^{3}~\mathrm{GeV}$, which corresponds to $H_{\mathrm{dom}} = 10^{-12}~\mathrm{GeV}$). For the computation we assume $g = 2 \pi /e$, $\mathcal{N}_{c, \mathrm{dom}} = 100$ and $B_0 = 10^{-15}~\mathrm{G}$.
Since now we are focusing on the case where the magnetic fields are generated before radiation domination, the bound of Eq.~\eqref{boundBefore} applies only for $H_i \geq H_{\mathrm{dom}}$. In the plot, the endpoints of the red and orange curves correspond to where $H_i = H_{\mathrm{dom}}$.

Also in this case, the exponential factor mainly sets the left-hand side of Eq.~\eqref{boundBefore}. Consequently, the bound reduces to the weak field condition up to an order-unity factor, as seen in the plot.
This is also seen by rewriting Eq.~\eqref{boundBefore} as,
\begin{equation}
    \mathscr{I} \gtrsim 1 + \frac{4}{g^2} \ln{\Tilde{x}_B} .
\end{equation}
The logarithmic factor can be generically neglected for an order-of-magnitude estimate of the bounds.

\subsection{Symmetry breaking scale and weak field condition}

For solitonic monopoles in spontaneously broken gauge theories, the symmetry breaking scale is typically of the order of $\mathrm{M} = m / g$. If the cosmic temperature or the Hubble rate upon magnetic field generation exceeds this scale, the symmetry is unbroken; then the monopole solution does not exist and the monopole mass limit of Eq.~\eqref{weakfieldMass} can be evaded. The magnetic field itself can also restore the symmetry if it exceeds the weak field limit of Eq.~\eqref{InstantonLimit} \cite{SS,KL,shore}. 

In parts of the region displayed in Figures~\ref{During} and~\ref{Before}, the temperatures $T_i$ and/or $T_{\mathrm{dom}}$ are larger than $\mathrm{M}$. However in such regions, the symmetry breaking after inflation can induce a monopole problem, which leads to further constraints on the monopole mass. We refer the reader to \cite{PrimMag} for a detailed discussion on this point.

In Figure~\ref{VelRatio}, with the choice of parameters there, the weak field condition implies $m \gtrsim 10^{14} \, \mathrm{GeV}$, and the condition of broken symmetry ($T_i, H_i < \mathrm{M}$) gives $m \gtrsim 10^{12} \, \mathrm{GeV}$. These conditions are violated in the plot by some of the curves with low masses. Such curves, hence, should be taken only as a qualitative indication of how light monopoles respond to primordial magnetic fields.

In Figure~\ref{Perri} we showed monopole bounds based on primordial magnetic fields during reheating, which were derived by evaluating the ratio $\Pi_{\mathrm{acc}}/\Pi_{\mathrm{red}}$ at the time $t_*$. (Recall that we do not need to specify the value of $H_i$ for these bounds, as long as $H_i > H_*$ is satisfied.) The magnetic field strength at $t_*$, i.e. $B_*$, satisfies the weak field condition in the entire range of $m$ and $T_{\mathrm{dom}}$ displayed in the plot. The condition $T_*, H_* < \mathrm{M}$ is also satisfied in the entire range, and thus the symmetry is broken at $t_*$.

We also remark that the constraint on the Hubble rate in Eq.~\eqref{eq:2.28} from the requirement that the magnetic field energy does not dominate the early universe is satisfied by the values of $H_i$ shown in Figures~\ref{VelRatio},~\ref{During},~\ref{Before}, and also by $H_*$ shown in Figure~\ref{Perri}.

\section{Conclusion}
\label{concl}

We carried out a comprehensive study of the monopole dynamics in the early universe and its back-reaction to the primordial magnetic fields from the time when the primordial magnetic fields have been generated to the epoch of $e^+ e^-$ annihilation. We then derived new bounds on the average abundance of magnetic monopoles in the universe by extending the Parker bound to the survival of the primordial magnetic fields.
From the analysis during radiation domination we re-derived the bound on the monopole number density given in \cite{Vachaspati}:
\begin{equation}
   n_0 \lesssim 10^{-21}~\mathrm{cm}^{-3} .
\end{equation}
Assuming the primordial magnetic fields to be generated prior to radiation domination, we obtained additional bounds by analyzing the magnetic field dissipation during the reheating period. 
For primordial magnetic fields produced at sufficiently early times (see Section~\ref{before3} for details), we derived:
\begin{equation}
      n_0 \lesssim 
    \begin{dcases}
    10^{-16}~\mathrm{cm}^{-3}  \left( \frac{B_0}{10^{-15}~\mathrm{G}} \right)^{3/5} \left( \frac{T_{\mathrm{dom}}}{10^{6}~\mathrm{GeV}} \right) \left( \frac{10}{g} \right)^{3/5} &,~~ m \ll \bar{m} , \\
    10^{-16}~\mathrm{cm}^{-3}  \left( \frac{m}{10^{14}~\mathrm{GeV}} \right) \left( \frac{T_{\mathrm{dom}}}{10^{6}~\mathrm{GeV}} \right) \left( \frac{10}{g} \right)^{2} &,~~ m \gg \bar{m} ,
    \end{dcases}
\end{equation}
where $\bar{m}$ is given by:
\begin{equation}
    \bar{m} \simeq 10^{14}~\mathrm{GeV}  \left( \frac{B_0}{10^{-15}~\mathrm{G}} \right)^{3/5} \left( \frac{g}{10} \right)^{7/5} \left( \frac{\mathcal{N}_{c, \mathrm{dom}}}{100} \right)^{2/5} .
\end{equation}
Here $T_{\mathrm{dom}}$ is the reheating temperature, $m$ is the monopole mass, $g$ is the magnetic charge of the monopoles, $B_0$ is the amplitude of the intergalactic magnetic field today, and $\mathcal{N}_{c}$ is the number of charged relativistic degrees of freedom.

In Fig.~\ref{Perri} we have shown the previous bounds on the monopole flux together with our results. 
For a sufficiently low reheating temperature, our bound is stronger than the original Parker bound and the limits from direct searches, even for GUT-scale monopoles. 
At low masses, our bound is stronger than that during radiation domination, Eq.~\eqref{PrimordialParker}, for $T_{\mathrm{dom}} \lesssim 10~\mathrm{GeV}$. 
In our analyses we assumed the plasma particles during the reheating epoch to always be in thermal equilibrium, however removing this assumption may further strengthen the bound on monopoles.

We also applied our bounds to monopoles that are Schwinger-produced in primordial magnetic fields, in order to obtain the most conservative condition for the survival of primordial magnetic fields.
We found that the bounds on the monopole density reduce to a weak field condition on the initial strength of the primordial magnetic fields,
\begin{equation}
\label{weakConclusion}
     B_i \lesssim \frac{4\pi m^{2}}{g^{3}} .
\end{equation}
This translates into a lower bound on the monopole mass shown in Eq.~\eqref{weakfieldMass}.
The work \cite{PrimMag} obtained a similar bound by only considering the acceleration of monopoles soon after they are pair produced. This indicates that the bound on the initial magnetic field strength does not improve significantly even when taking into account the integrated effect of the monopole acceleration over the entire cosmological history. This insensitivity to the detailed dynamics of the monopoles is due to the exponential dependence of the monopole production rate on the magnetic field strength.
We thus conclude that as long as the initial amplitude of the primordial magnetic field is sufficiently below the bound of Eq.~\eqref{weakConclusion}, the back-reaction from Schwinger-produced monopoles can be safely ignored.

\acknowledgments

T.K. acknowledges support from the INFN program on Theoretical Astroparticle Physics, and JSPS KAKENHI (Grant Numbers JP16H06492, JP22K03595). The authors thank Hiroyuki Tashiro for helpful discussions.

\appendix
\section{The universe from the end of inflation to the matter-radiation equality}
\label{app1}

In this appendix, we derive the relations between the Hubble rate, the cosmic temperature and the scale factor both during reheating and in the following radiation-dominated epoch.
We define $t_{\mathrm{end}}$ as the time of the end of inflation, when the universe begins to be dominated by an oscillating inflaton field. 
The inflaton eventually decays into radiaton, and at time $t_{\mathrm{dom}}$ the radiation component starts to dominate the universe. Then at time $t_{\mathrm{eq}}$, the time of matter-radiation equality, matter domination begins. Setting $t_{\mathrm{end}} = t_{\mathrm{dom}}$ corresponds to the case of an instantaneous reheating.

During radiation domination, for times $t_{\mathrm{dom}} \ll t \ll t_{\mathrm{eq}}$, the Friedmann equation gives $3M^2_{\mathrm{Pl}} H^2 \simeq \rho_{\mathrm{rad}}$, with $\rho_{\mathrm{rad}} = (\pi^2/30)g_* T^4$ the radiation energy density and $T$ the radiation temperature. Considering the expression for the entropy density $s = (2 \pi^2/45) g_{*s} T^3$, we get (the subscript ``$0$'' denotes quantities in the present universe):
\begin{equation}
\label{HradAppendix}
    H \simeq \left(\frac{45}{128 \pi^{2}}\right)^{1 / 6} \frac{g_{*}^{1 / 2}}{g_{*s}^{2 / 3}} \frac{s_{0}^{2 / 3}}{M_{\mathrm{Pl}}}\left(\frac{a_{0}}{a}\right)^{2}, \quad T \simeq \left(\frac{45}{2 \pi^{2}} \frac{s_{0}}{g_{*s}}\right)^{1 / 3} \frac{a_{0}}{a} .
\end{equation}
Here we have assumed the conservation of entropy until today, i.e. $s \propto a^{-3}$. The relation between the scale factor and the temperature during radiation domination is as follows:
\begin{equation}
\label{gstar1}
    \left( \frac{a_0}{a} \right)^3 = \frac{g_{* s} T^3}{g_{* s, 0} T_0^3} .
\end{equation}

During reheating, for times $t_{\mathrm{end}} \ll t \ll t_{\mathrm{dom}}$, the oscillating inflaton field decays perturbatively into relativistic particles. During this period the universe is effectively matter-dominated, with $H \propto a^{-3/2}$. 
Let us assume that the relativistic particles are in thermal equilibrium also during this period, so that $\rho_{\mathrm{rad}} = (\pi^2 / 30) g_* T^4$ holds. For simplicity, we ignore the time dependence of $g_{*(s)}$ before radiation domination. This amounts to assuming that no additional relativistic degrees of freedom appear as one goes back in time from $t = t_{\mathrm{dom}}$. Then the radiation density, which is sourced by the inflaton decay, evolves as $\rho_{\mathrm{rad}} \propto a^{-3/2}$.
The scaling behavior of $\rho_{\mathrm{rad}}$ can be verified by solving the continuity equation $\dot{\rho}_{\mathrm{rad}} + 4H \rho_{\mathrm{rad}} = \Gamma_{\phi} \rho_{\phi}$, where $\Gamma_{\phi}$ is the inflaton decay rate and $\rho_{\phi} = \rho_{\phi \mathrm{end}} (a_{\mathrm{end}}/a)^3 e^{-\Gamma_{\phi} (t - t_{\mathrm{end}})}$ is the energy density of the inflaton field \cite{KolbTurner}. The radiation temperature thus redshifts as $T \propto a^{-3/8}$. We can also write the relation between the scale factor and the temperature before radiation domination as:
\begin{equation}
\label{agstar}
    \frac{a_0}{a} = \frac{a_{\mathrm{dom}}}{a} \frac{a_0}{a_{\mathrm{dom}}} = \left( \frac{T}{T_{\mathrm{dom}}} \right)^{8/3} \left( \frac{g_{*s,\mathrm{dom}}}{g_{*s,0}} \right)^{1/3} \left(\frac{T_{\mathrm{dom}}}{T_0}\right) .
\end{equation}

We summarize the dependence on the scale factor of the Hubble rate and of the temperature in the following expressions:
\begin{equation}
\label{HTevolution}
\begin{aligned}
    H \simeq H_{\mathrm{dom}} & \min \left\{
    \left(\frac{a_{\mathrm{dom}}}{a}\right)^{3 / 2}, \left(\frac{g_*}{g_{*,\mathrm{dom}}} \right)^{1/2} \left(\frac{g_{*s,\mathrm{dom}}}{g_{*s}} \right)^{2/3} \left(\frac{a_{\mathrm{dom}}}{a}\right)^{2} \right\} ,\\
    T & \simeq T_{\mathrm{dom}} \min \left\{ \left(\frac{a_{\mathrm{dom}}}{a}\right)^{3 / 8}, \left( \frac{g_{*s,\mathrm{dom}}}{g_{*s}} \right)^{1/3} \left(\frac{a_{\mathrm{dom}}}{a}\right) \right\} .
\end{aligned}
\end{equation}
The first expression in the curly brackets holds for $t_{\mathrm{end}} < t < t_{\mathrm{dom}}$ and the second for $t_{\mathrm{dom}} < t < t_{\mathrm{eq}}$. 

By extrapolating Eq.~\eqref{HradAppendix} to the time when radiation domination begins, we can obtain the relations between $H_{\mathrm{dom}}$, $T_{\mathrm{dom}}$, and $a_{\mathrm{dom}}$.
After substituting numerical values for the reduced Planck mass $M_{\mathrm{Pl}}$ and for the cosmological parameters we get:
\begin{equation}
\label{aTdom}
    \frac{a_{0}}{a_{\mathrm{dom}}} \simeq 10^{29}\left(\frac{H_{\mathrm{dom}}}{10^{14} \mathrm{GeV}}\right)^{1/2}, \quad T_{\mathrm{dom}} \simeq 10^{16} \mathrm{GeV}\left(\frac{H_{\mathrm{dom}}}{10^{14} \mathrm{GeV}}\right)^{1/2} .
\end{equation}
We also underline that these expressions present only a weak dependence on $g_{*(s)}$ and then the order-of-magnitude estimates are not affected by its precise value. 
Reversing the first line in Eq.~\eqref{HTevolution} and ignoring the contributions of $g_{* (s)}$, we express the scale factor as a function of the Hubble rate:
\begin{equation}
\label{aidom}
    \frac{a_{\mathrm{dom}}}{a} \simeq \left( \frac{H}{H_{\mathrm{dom}}} \right)^{1/2}  \max \left \{ \left( \frac{H}{H_{\mathrm{dom}}} \right)^{1/6}\ ,\ 1 \right \} .
\end{equation}
The first term inside the curly brackets holds for $t_{\mathrm{end}} < t < t_{\mathrm{dom}}$ and the second for $t_{\mathrm{dom}} < t < t_{\mathrm{eq}}$.

\end{document}